\documentclass[12pt]{article}
\pdfoutput=1
\usepackage{tikz}
\usetikzlibrary{matrix,arrows,positioning,shapes,snakes,fadings,decorations.pathmorphing,decorations.pathreplacing,automata,shadows}
\usepackage[thicklines]{cancel}
\usetikzlibrary{calc,shapes.callouts,shapes.arrows}
\usepackage{jheppub}
\usepackage{amsmath,amssymb,euscript,array,mathrsfs,appendix,ctable,marvosym,calc,array}
\usepackage{arydshln}
\usepackage{todonotes}
\usepackage{graphicx}
\usepackage[normalem]{ulem}

\usepackage{blkarray}
\usepackage[normalem]{ulem}

\newcommand*\circled[1]{\footnotesize\tikz[baseline=(char.base)]{%
            \node[shape=circle,fill=black!20,draw,inner sep=2pt] (char) {#1};}}

\usepackage{enumitem}

\newcolumntype{C}[1]{>{\centering\arraybackslash$}m{#1}<{$}}
\newlength{\mycolwd}                                         
\newcommand{\JLMc}[1] {{\color{red} #1}}

\newcommand\blank[1]{#1}
\renewcommand\blank[1]{}
\def\Buildrel#1\over#2\under#3{\mathrel{\mathop{\kern0pt
#2}\limits^{#1}_{#3}}}

\usepackage{color}
\definecolor{lightgray}{gray}{0.75}

\def\AA{{\mathfrak L}}

\def\CF{{\cal F}}
\def\PP{{\mathbb P}}

\def\JJ{\mathscr{J}}

\def\mpsu{\mathfrak{psu}}

\def\SO{\text{SO}}

\def\msl{\mathfrak {sl}}

\def\mf{\mathfrak f}

\def\msu{\mathfrak{su}}

\def\msl{\mathfrak{sl}}

\def\PSU{\text{PSU}}

\def\p{\pi}

\newcommand{\Tr}{\operatorname{Tr}}
\newcommand{\STr}{\operatorname{STr}}
\newcommand{\IM}{\operatorname{Im}}
\newcommand{\RE}{\operatorname{Re}}

\def\B0{{\boldsymbol 0}}
\def\BF{{\boldsymbol F}}

\def\Bvarpi{{\boldsymbol\varpi}}

\def\CP{{\mathbb C}P}

\def\sst{\scriptscriptstyle}
\def\det{{\rm det}}
\def\SU{\text{SU}}
\def\Sp{\text{Sp}}

\def\Dbarslash{\,\,{\raise.15ex\hbox{/}\mkern-12mu {\bar D}}}
\def\Dslash{\,\,{\raise.15ex\hbox{/}\mkern-12mu D}}
\def\delslash{\,\,{\raise.15ex\hbox{/}\mkern-9mu \partial}}
\def\delbarslash{\,\,{\raise.15ex\hbox{/}\mkern-9mu {\bar\partial}}}

\def\ket#1{| #1\rangle}

\newcommand{\MAT}[1]{\begin{pmatrix} #1\end{pmatrix}}

\newcommand{\EQ}[1]{\begin{equation}\begin{split} #1
\end{split}\end{equation}}

\newcommand{\FIG}[1]{\begin{figure}[ht]\begin{center} #1 \end{center}\end{figure}}

\title{Giant Magnons of String Theory in the Lambda Background}
\author[a]{Calan Appadu,}
\author[a]{Timothy J. Hollowood,}
\author[b]{J. Luis Miramontes,}
\author[a]{Dafydd Price,}
\author[c]{and David M. Schmidtt} 
\affiliation[a]{Department of Physics, Swansea University, Swansea, SA2 8PP, U.K.}
\affiliation[b]{Departamento de F\'\i sica de Part\'\i culas and IGFAE,
Universidad de Santiago de Compostela, 15782 Santiago de Compostela, Spain}
\affiliation[c]{Departamento de F\'\i sica, Universidade Federal de S\~ao Carlos, \\
Caixa Postal 676, CEP 13565-905, S\~ao Carlos-SP, Brazil}

\emailAdd{t.hollowood@swansea.ac.uk}
\emailAdd{jluis.miramontes@usc.es}
\emailAdd{david@df.ufscar.br}

\abstract{The analogues of  giant magnon configurations are studied on the string world sheet in the lambda background. This is a discrete deformation of the AdS$_ 5{\times} S^5$ background that preserves the integrability of the world sheet theory. Giant magnon solutions are generated using the dressing method and their dispersion relation is found. This reduces to the usual dyonic giant magnon dispersion relation in the appropriate limit and becomes relativistic in another limit where the lambda model becomes the generalized sine-Gordon theory of the Pohlmeyer reduction. The scattering of giant magnons is then shown in the semi-classical limit to be described by the quantum S-matrix that is a quantum group deformation of the conventional giant magnon S-matrix. It is further shown that in the small $g$ limit, a sector of the S-matrix is related to the XXZ spin chain whose spectrum matches the spectrum of magnon bound states.}

\notoc
\begin{document}

\maketitle

\newpage

\section{Introduction}\label{s1}

In this paper, we study excitations on the Green-Schwarz world sheet of the string in the lambda background that generalize the giant magnons of the string in AdS$_ 5{\times} S^5$. The lambda background can be thought of as a discrete deformation of the non abelian T-dual of the string in  AdS$_ 5{\times} S^5$ with respect to the full super group $\PSU(2,2|4)$ symmetry \cite{Hollowood:2014qma}, generalizing an original idea for bosonic sigma models \cite{Sfetsos:2013wia} (see also the earlier \cite{Rajeev:1988hq,Balog:1993es,Sfetsos:1994vz,Evans:1994hi,Polychronakos:2010hd} and the more recent \cite{Hollowood:2014rla}). The fact that it is a consistent background for the string has been investigated in \cite{Sfetsos:2014cea,Demulder:2015lva,Borsato:2016zcf,Chervonyi:2016ajp,Borsato:2016ose}, with explicit results for AdS$_n\times S^n$ with $n=2$~\cite{Borsato:2016zcf} and 3~\cite{Chervonyi:2016ajp} and on general grounds for $n=5$ in \cite{Borsato:2016ose}. 
There is large literature on various kinds of integrable deformations of the AdS$_ 5{\times} S^5$ string theory. 
Works which specifically investigate the {\it lambda deformation\/} of string theory include \cite{Arutyunov:2012zt,Arutyunov:2012ai,Delduc:2013fga,vanTongeren:2013gva,Hoare:2014pna,Vicedo:2015pna,Hoare:2015gda,Sfetsos:2015nya,Hollowood:2015dpa,Klimcik:2016rov,Chervonyi:2016bfl,Schmidtt:2016tkx,Schmidtt:2017CS,Appadu:2017fff}.\footnote{Note that the terminology {\it lambda deformation\/} seems to have become established even though the deformation parameter of the string theory is really the integer $k$. The couplings $\lambda$ and $k$ are related in \eqref{crel}. In the semi-classical limit $k\to\infty$ and then $\lambda$ labels a family of inequivalent semi-classical theories.
}

Specifically in this paper we:
\begin{enumerate}[label=\protect\circled{\arabic*}]
\item construct the giant magnons in the lambda model using the dressing method.
\item calculate their charges, including the energy and momentum thereby establishing their dispersion relation.
\item analyse the scattering of giant magnons at the classical level extracting the time delays.
\item match the spectrum of giant magnons at the quantum level with short, atypical or BPS representations of the underlying Lie super algebra and thereby show that the dispersion relation is exact at the semi-classical level.
\item show that the exact S-matrix constructed in \cite{Hoare:2011nd,Hoare:2011wr,Hoare:2012fc} matches the classical scattering of the magnons in the semi-classical limit using the Jackiw-Woo formula. 
\item solve a puzzle posed in \cite{Hoare:2012fc} about the nature of the bound-states poles and whether the bound states are associated to the AdS$_5$ or $S^5$ part of the geometry: The answer is always the $S^5$ part.
\item we show that in the limit $\lambda\to1$ (i.e.~$g\to0$ where $g$ is defined in \eqref{crel}) a sub-sector of the magnons have a spectrum and a scattering theory which matches the XXZ spin chain \cite{vanTongeren:2013gva}. This provides some clues as to how the lambda deformation can be interpreted at the level of the ${\cal N}=4$ theory.
\item appendices \ref{a1}-\ref{a3} contain more detailed analyses of various aspects of the lambda model: conserved charges, Noether symmetries and symplectic form.
\end{enumerate}

The simplest way to construct the lambda model (the Green-Schwarz world sheet theory of the string in the lambda background) is to write the Green-Schwarz sigma model for the string in AdS$_ 5{\times} S^5$ \cite{Metsaev:1998it} in first order form. In the present work we will work in conformal gauge $\gamma_{\mu\nu}=e^\phi\eta_{\mu\nu}$ for simplicity, so that\footnote{The superscripts denote the grade of an element of the Lie super algebra $\mpsu(2,2|4)$ under the $\mathbb Z_4$ automorphism that underlies the semi-symmetric space $F/G=\PSU(2,2|4)/\Sp(2,2){\times}\Sp(4)$.}
\EQ{
S_\sigma=-4g\int d^2x\,\STr\Big[A_+^{(2)}A^{(2)}_-+\frac12A^{(1)}_+A^{(3)}_--\frac12A^{(1)}_-A^{(3)}_++\nu F_{+-}\Big]\ ,
\label{psr}
}
where $F_{\mu\nu}$ is the field strength of the $\PSU(2,2|4)$ gauge field $A_\mu$. The Lie super algebra-valued field $\nu$ acts as a Lagrange multiplier that imposes the condition that there exists a group valued field $f$ such that $A_\mu=f^{-1}\partial_\mu f$, so that $S_\sigma$ becomes the action of the $\text{AdS}_ 5{\times} S^5$ string sigma model~\cite{Metsaev:1998it}. Alternatively, if we integrate the gauge field $A_\mu$ out, $S_\sigma$ specifies the non abelian T-dual of the string in $\text{AdS}_ 5{\times} S^5$ with respect to the full supergroup $\PSU(2,2|4)$ symmetry.

Inspired by the strategy of \cite{Sfetsos:2013wia} for bosonic sigma models, the lambda model is obtained by enhancing $\nu$ to a $F=\PSU(2,2|4)$ group valued field $\CF$ and replacing the Lagrange multiplier coupling in \eqref{psr} with the gauged WZW model \cite{Hollowood:2014qma} 
\EQ{
-4g\int d^2x\,\STr\big(\nu F_{+-}\big)\longrightarrow S_\text{gWZW}[\CF,A_\mu]\ .
}
The current of the sigma model $A_\mu$ now becomes re-interpreted as the gauge field for the  gauged WZW theory for the supergroup $F$ gauged with respect to the anomaly free vector subgroup $F_V\subset F_L\times F_R$. In addition, in order to obtain an integrable theory on the world sheet, the terms in the original sigma model action must also be suitably deformed,
\EQ{
S_\lambda=S_\text{gWZW}[\CF ,A_\mu]-&\frac {k}{\pi}\int d^2x\,\STr\,\big[
A_+(\Omega_+-1)A_-\big]\ ,
\label{dWZW}
}
where
\EQ{
\Omega_\pm=\PP^{(0)}+\lambda^{\pm1}\mathbb P^{(1)}+\lambda^{-2}\mathbb P^{(2)}+\lambda^{\mp1}\mathbb P^{(3)}\ .
\label{dom}
}
In the above, $\mathbb P^{(i)}$ are projectors onto the eigenspaces of the Lie super algebra $\mf=\oplus_{i=0}^3\mf^{(i)}$ under the $\mathbb Z_4$ automorphism, and $\STr(A_+ \Omega_+ A_-)= \STr(\Omega_- A_+  A_-)$. The lambda model has two coupling constants. The first one is $k\in {\mathbb Z}$, which is the 
usual quantized level of the super WZW part of the action. 
The second is $\lambda\in (0,1)$, which parameterizes the deformation and is marginal at least to one loop~\cite{Appadu:2015nfa}.\footnote{It is also marginal to one loop for the hybrid formalism of the superstring~\cite{Schmidtt:2016tkx}.} The 
action~\eqref{psr}
is then recovered---at least heuristically---in the joint limit $k\to\infty$ and $\lambda\to1$ with  $4\pi g=k(\lambda^{-2}-1)$ fixed and with $\CF$ expanded around the identity with  $\CF=\exp(4\pi g\nu/k)$.\footnote{The large $k$ limit yields the non abelian T dual of the original sigma model. The extent to which the large $k$ limit actually yields the original $\text{AdS}_ 5{\times} S^5$ string theory rests on whether non abelian T duality, which at the classical level is known to be a canonical transformation \cite{Lozano:1995jx}, becomes a fully fledged quantum equivalence. There is evidence from the S-matrix that makes this plausible. In the limit $k\to\infty$, the lambda model S magnon matrix becomes equal to the original magnon S-matrix of the sigma model up to an IRF-to-vertex transformation which one can think of as a change of basis in the Hilbert space \cite{Hoare:2013hhh} (this is also true on bosonic lambda models \cite{Appadu:2017fff}).} This suggests the following convenient parameterization of the coupling $\lambda$  in terms of $k$ and $g$
\EQ{
\frac1{\lambda^2}=1+\frac{4\pi g}k\ .
\label{crel}
}
The ${\cal O}(A^2)$ term in the action \eqref{dWZW} actually breaks the vector $F_V=\PSU(2,2|4)$ gauge symmetry down to the bosonic subgroup $G_V=\Sp(2,2){\times}\Sp(4)(\simeq\SO(1,4){\times}\SO(5))$.  Importantly, however, a deformation of part of the fermionic component of $F_V$ gives rise to a set of kappa symmetries that are needed for a consistent Green-Schwarz sigma model \cite{Hollowood:2014qma,usint}. 

The fields $A_\mu$ can be integrated out which amounts to imposing their equations of motion, which take the form of the constraints
\EQ{
\CF ^{-1}\partial_+\CF +\CF ^{-1}A_+\CF =\Omega_-A_+\ ,\qquad -\partial_-\CF\CF^{-1} +\CF A_-\CF^{-1} =\Omega_+A_-\ .
\label{rup}
}
These constraints are partly second class and partly first class to reflect the remaining gauge and kappa symmetries \cite{Hollowood:2014qma}.

The resulting sigma model for the group field $\CF$ is complicated and appears to have no obvious symmetries. However, looks can be deceiving: there are actually conventional Noether symmetries that we will exploit for gauge fixing the world sheet theory. However, these world sheet symmetries do not lift as to isometries of the space time; indeed, the lambda model background obtained by integrating out the auxiliary field $A_\mu$ does not have any Killing vectors \cite{Sfetsos:2014cea}. In addition, at the quantum level integrating out $A_\mu$ produces a super determinant that gives rise to a dilaton on the world sheet.
For the simpler lambda models corresponding to  $\text{AdS}_2\times S^2$ and $\text{AdS}_3\times S^3$, it has been explicitly checked that the deformed sigma model, including this dilaton, is Weyl invariant and, therefore, specify a consistent superstring background~\cite{Borsato:2016zcf,Chervonyi:2016ajp}. Recently, the $\text{AdS}_ 5{\times} S^5$ lambda model has been shown to behave exactly in the same way~\cite{Borsato:2016ose}. 

Like the sigma model, the lambda model is an integrable theory. This is uncovered by writing the equations of motion in Lax form
\EQ{
[\partial_\mu+{\AA}_\mu(z),\partial_\nu+{\AA}_\nu(z)]=0\ ,
\label{leq2}
}
with a spectral parameter $z$. The constraints \eqref{rup}, which are the equations of motion of $A_\mu$, can be used to write the Lax connection purely in terms of $A_\mu$.
Then, the constraints~\eqref{rup} allow one to reconstruct the group field $\CF$ from the ``wave function" $\Psi$ of the associated linear system
\EQ{
\big(\partial_\mu+\AA_\mu(x;z)\big)\Psi(x;z)=0\ ,
\label{lsy}
}
as
\EQ{
\CF(x)=\Psi(x;\lambda^{1/2})\Psi^{-1}(x;\lambda^{-1/2})\ .
\label{eyy}
}

The zero-curvature condition~\eqref{leq2} involves the components of
\EQ{
J_\pm= A_\pm^{(0)} + \lambda^{\mp1/2} A_\pm^{(1)} + \lambda^{-1}A_\pm^{(2)}+\lambda^{\pm1/2} A_\pm^{(3)}\ ,
}
so that
\EQ{
{\AA}_\pm(z)=J_\pm^{(0)}+z J_\pm^{(1)}+z^{\pm2} J_\pm^{(2)}+z^{-1} J_\pm^{(3)}
\label{mm2b}
}
and $J_\pm=\AA_\pm(1)$. Written in terms of $J_\mu$,~\eqref{leq2} is the zero-curvature condition satisfied by the current of the Green-Schwarz sigma model where $J_\mu$ is defined in terms of a group valued field 
$f\in\text{PSU}(2,2|4)$ by means of
\EQ{
J_\mu = f^{-1}\partial_\mu f\ .
\label{defJ}
}
It is a crucial fact for us that the sigma and lambda models share the same linear system, so that a solution $\Psi(x;z)$ of it provides simultaneously a solution of  both the equations of motion of  the sigma and the lambda model for generic values of $\lambda$.
In particular, the sigma model field is extracted from the wave function via
\EQ{
f=\Psi^{-1} (x;1)\ .
\label{eyy2}
}
Following the standard lore of integrable systems, e.g.~\cite{BBT}, it is useful to recall that $\Psi$ is defined up to a right multiplication by a space-time independent group element, and that this freedom can be fixed by imposing a normalization condition like $\Psi(0;z)=1$, which is satisfied by the vacuum solution $\Psi_0$ given by \eqref{vlsy}. However, the freedom to change this, corresponds to generalizing  eqs.~\eqref{eyy} and~\eqref{eyy2} to
\EQ{
\CF(x)=\Psi(x;\lambda^{1/2})V(\lambda) \Psi^{-1}(x;\lambda^{-1/2})\,,\qquad
f=V\Psi^{-1} (x;1)\ .
\label{eyy3}
}
respectively, where $V(\lambda)$ and $V\equiv V(1)$ are space-time independent elements of $F$. The freedom to choose $V$ is a reflection of the well know invariance of the sigma model under global $F_L$ transformations $f\to Vf$, with $V\in F$. Similarly, the freedom to choose $V(\lambda)$ implies a global hidden Noether symmetry of the lambda model that we exploit later for gauge fixing. We describe the symmetries in Appendix \ref{a2}.

Notice that~\eqref{rup} and~\eqref{leq2} summarize the equations of motion of the fields $\CF$ and $A_\mu$, but we still have to impose the equations of motion of the world sheet metric: the Virasoro constraints. On shell and in conformal gauge, they read
\EQ{
\STr(J_+^{(2)}J_+^{(2)})=\STr(J_-^{(2)}J_-^{(2)})=0\ .
\label{vc}
}

In the present work, we will investigate how soliton excitations on the sigma model world sheet known as giant magnons appear in the lambda model. In the AdS$_ 5{\times} S^5$ string, these excitations are a key ingredient in the relation between the world sheet theory and the dual ${\cal N}=4$ gauge theory. This becomes particularly apparent in the Hofman-Maldacena (HM) limit \cite{Hofman:2006xt} where the world sheet theory that describes a closed string effectively becomes decompactified. In this limit it makes sense to define asymptotic in and out states and a conventional S-matrix. Because the theory is integrable this S-matrix is factorizable. However, in order to gauge fix the world sheet theory one works in a light cone gauge which can be interpreted as a special parametrization of the degrees of freedom around a particular ``vacuum" solution; in this case the 
BMN solution \cite{Berenstein:2003gb}. This solution describes a string that is compressed to a point traversing a null geodesic of AdS$_ 5{\times} S^5$ that involves an orbit of the equator of $S^5$. The physical subspace corresponds to certain transverse degrees of freedom to this vacuum solution. 

The integrable scattering theory is rather unconventional because the gauge fixed theory is not Lorentz invariant. This is highlighted by the dispersion relation for the giant magnons:
\EQ{
{\cal E}^2={\cal Q}^2+16g^2\sin^2\Big[\frac{\cal P}{4g}\Big]\ .
\label{yr2}
}
In the above, ${\cal Q}=1,2\ldots$ is a discrete charge that takes the basic giant magnon ${\cal Q}=1$ solution to its dyonic generalization that carries an abelian charge under the symmetry left unbroken by the vacuum solution \cite{Dorey:2006dq,Chen:2006gea}.

On the dual gauge theory side, the HM limit corresponds to focussing on operators that are built on the dual to the BMN solution, which is a long operator formed from one of the scalar fields of the ${\cal N}=4$ theory $\Tr(X^L)$, $L\to\infty$. Magnons on the world sheet correspond to excitations of the associated spin chain that build up a spectrum of operators around the BMN vacuum. For excitations in the so-called $\msu(2)$ sector, the spin chain is precisely the Heisenberg XXX spin chain describing single trace operators that involve products of infinitely many $X$'s and a finite number of $Y$, one of the other complex scalar fields. The story is long and fascinating (reviewed in the series of articles \cite{Beisert:2010jr}) and we touch on it and the relation to the lambda theory in section \ref{s6}.

In this work, we ask: what are the analogues of the giant magnons of the lambda theory?
Since the equations of motion of the lambda theory and the sigma model involve the same linear system it seems obvious that the giant magnons of the sigma model are directly related to soliton solutions of the lambda theories. We show that this intuition is correct. In particular, the dispersion relation \eqref{yr2} generalizes to 
\EQ{
\boxed{(\lambda^{-1}+\lambda)^2\sin^2\left[\frac{\pi{\cal E}}{k(\lambda^{-1}+\lambda)}\right]-(\lambda^{-1}-\lambda)^2\sin^2\left[\frac{\pi{\cal P}}{k(\lambda^{-1}-\lambda)}\right]=4\sin^2\left[\frac{\pi{\cal Q}}{2k}\right]\ .}
\label{dis}
}
This dispersion relation has already appeared in a different parameterization in \cite{Hoare:2011wr,Hoare:2012fc,Hoare:2013hhh,Hollowood:2014fha}.
The ordinary dyonic giant magnon dispersion relation \eqref{yr2} is obtained by taking $k\to\infty$, with $g$ fixed, i.e.~$\lambda\to1$ in~\eqref{crel}. We call this the ``sigma model limit".

The dispersion relation above is quite remarkable. Not only does it yield \eqref{yr2} in the sigma model limit but it becomes relativistic in another interesting limit $g\to\infty$ with $k$ fixed, i.e.~$\lambda\to0$. We call this the 
 ``sine-Gordon limit":
\EQ{
{\cal E}^2-{\cal P}^2=\Big(\frac{2k}\pi\sin\frac{\pi{\cal Q}}{2k}\Big)^2\ .
\label{mss}
} 
This reflects the fact that in this limit the gauge fixed world sheet theory becomes a relativistic QFT identifed as a generalized sine-Gordon theory \cite{Grigoriev:2007bu,Mikhailov:2007xr,Hollowood:2011fq,susy-flows} that is associated to the Pohlmeyer reduction of the sigma model  \cite{Miramontes:2008wt}.

The giant magnons of the AdS$_ 5{\times} S^5$ sigma model have an exact quantum S-matrix whose symmetry involves the Yangian of the subgroup of $\PSU(2,2|4)$ left invariant by the BMN solution. At the algebra level, this is a centrally extended version of $\mpsu(2|2){\oplus}\mpsu(2|2)$ \cite{Beisert:2005tm,Beisert:2010jr}.

One naturally wonders about the scattering of giant magnons of the lambda model. There is a natural candidate for their S-matrix that was constructed and analysed in \cite{Hoare:2011nd,Hoare:2011wr,Hoare:2012fc,Hoare:2013hhh}. This S-matrix involves a quantum group deformation of the Yangian invariant AdS$_ 5{\times} S^5$ S-matrix with a deformation parameter $q=\exp(i\pi/k)$. The states of this S-matrix theory are kinks that transform in the Interaction Round a Face (IRF), or Restricted Solid On Solid (RSOS), form of the quantum group that naturally describes the representation theory with $q$ a root of unity. In this paper, we will show that this scattering theory consistently  describes the scattering of giant magnons in the lambda model by subjecting it to a semi-classical test based on the Jackiw-Woo formula \cite{Jackiw:1975im} which relates the S-matrix to the time delay experienced as magnons classically scatter.
 
Finally we ask the question: what can the scattering theory tell us about what could be the dual theory to the lambda model? A more modest question is to ask in the limit $g\to0$ and in the $\msu(2)$ sector, what deformation of the XXX spin chain describes the spectrum of the giant magnons in the lambda theory. Perhaps not surprisingly, given the form of the S-matrix, it turns out to be the XXZ spin chain \cite{vanTongeren:2013gva}. This suggests that the dual theory is the quantum group deformed version of the quantum mechanical matrix model describing the ${\cal N}=4$ theory on $S^3$ suggested in \cite{Berenstein:2004ys}. The additional novel element is the fact that $q$ is a root of unity and the quantum group is realized in its IRF/RSOS form.
 
This paper is organized as follows: in section \ref{s10} we discuss the spectrum of states described by the factorizable scattering theory of \cite{Hoare:2011wr,Hoare:2012fc,Hoare:2013hhh,Hollowood:2014fha}. This scattering theory describes a quantum group deformation of the S-matrix of the giant magnons of the $\text{AdS}_ 5{\times} S^5$ theory. The states are described by the dispersion relation \eqref{dis}.
In this section, we show that the excitations come in two distinct branches, ``magnon" and ``soliton", distinguished by the value of the momentum. This is something that
could not be realized in a relativistic theory where a state with non vanishing momentum is just a boost of the same state at rest.
We also propose a solution to the puzzle posed in \cite{Hoare:2012fc} about the nature of the bound-states poles.
In section \ref{s2} we describe how the world sheet lambda model can be gauge fixed and how one defines the analogue of the Hofman-Maldacena limit where the world sheet effectively decompactifes and one can talk about asymptotic states and an S-matrix. In section \ref{s3}, we construct the magnon solutions explicitly via the dressing method. We then analyse the solutions and extract their charges and show that their dispersion relation is \eqref{dis}. This shows that \eqref{dis} does not receive quantum corrections, up to possible shifts in the level $k$, of the kind that may be similar to the level shift in WZW models. 
We also show that the giant magnon solutions can be naturally understood as kinks.
In section \ref{s3.3}, we consider the classical scattering of magnons and derive a formula for the time delay experienced by a magnon as it passes through another magnon.
In section \ref{s5} we turn to the quantum magnons and their S-matrix. Section  \ref{s5.3} describes the exact quantum S-matrix of the magnons and a particular sub sector of their bound states whose S-matrix can be determined by using the bootstrap equations. In section \ref{s5.4}, we take the semi-classical limit of the S-matrix elements for magnon bound states with large charge. The technical difficulties here are dealing with the dressing phase. This leads to a detailed semi-classical comparison of the quantum S-matrix with the classical time delays: we find perfect consistency. In section \ref{s6}, we consider the $g\to0$ limit of the magnons and their S-matrix and find that there is a relation with the XXZ Heisenberg spin chain that generalizes the connection of the AdS$_ 5{\times} S^5$ magnons with the the XXX spin chain. The deformation parameter that takes XXX to XXZ is $\Delta=\cos(\pi/k)$. More detailed properties of the lambda model are analysed in appendices \ref{a1}-\ref{a3}.
 
\section{The quantum spectrum}\label{s10}

Before we analyse the classical world sheet theory, it is worth pausing to consider the quantum spectrum and S-matrix constructed in \cite{Hoare:2011nd,Hoare:2011wr,Hoare:2012fc,Hoare:2013hhh}. It is the main hypothesis of this work that this S-matrix theory describes the spectrum and scattering of magnons in the lambda string theory. We shall show that there is a subtlety in defining the physical energy in the S-matrix theory that when properly understood solves the puzzle posed in \cite{Hoare:2012fc} about the nature of bound-states poles, and a specific quantization rule for the charge carried by the magnons.

\vspace{0.5cm}

States transform in representations of the quantum group version of the symmetry left invariant  by the BMN  state. This is a product of two copies of a quantum super group. The appropriate representations are special short, or BPS, or atypical, representations of a central extension of the corresponding quantum deformed Lie super algebra $U_q(\msu(2|2))$.
In the undeformed (sigma model) limit, $q\to1$, each of the groups $\SU(2|2)$ has a bosonic subgroup $\SU(2){\times}\SU(2)$ that is related to isometries of the $\text{AdS}_ 5{\times} S^5$ background in the way illustrated in \ref{f1}.

\FIG{
\begin{tikzpicture} [scale=0.9]
\node at (-1,0) (a1) {$S^5$};
\node at (1.7,0) (a2) {$\SU(4)$};
\node[blue] at (4,0.8) (a3) {$\SU(2)_3$};
\node[blue] at (4,-0.8) (a4) {$\SU(2)_4$};
\node at (7,0) (a5) {$\SU(2|2)$};
\draw[-] (a1) -- (a2);
\draw[-] (a2) -- (a3);
\draw[-] (a2) -- (a4);
\begin{scope}[yshift=3cm]
\node at (-1,0) (b1) {AdS$_5$};
\node at (1.7,0) (b2) {$\SU(2,2)$};
\node[red] at (4,0.8) (b3) {$\SU(2)_1$};
\node[red] at (4,-0.8) (b4) {$\SU(2)_2$};
\node at (7,0) (b5) {$\SU(2|2)$};
\draw[-] (b1) -- (b2);
\draw[-] (b2) -- (b3);
\draw[-] (b2) -- (b4);
\end{scope}
\draw[-] (4.8,3.8) -- (5,3.8) -- (5,0.8) -- (4.8,0.8);
\draw[-] (4.8,2.2) -- (5.2,2.2) -- (5.2,-0.8) -- (4.8,-0.8);
\draw[-] (5.2,0) -- (6,0);
\draw[-] (5,3) -- (6,3);
\draw[-] (a5) -- (8.6,0);
\draw[-] (b5) -- (8.6,3);
\node at (10,0) {${\footnotesize \left(\begin{array}{cc|cc} \,0\, &\, 0\, & \,0\,& \,0\,\\ \,0\, & \color{red}{*}& \,0\,& *\,\\ \hline 0&0&0&0\\ 0&*&0&\color{blue}{*}\end{array}\right)}$};
\node at (10,3) {${\footnotesize\left(\begin{array}{cc|cc} \color{red}{*} & 0 & *& 0\\ \,0\, & \,0\, & \,0\,& \,0\,\\ \hline *&0&\color{blue}{*}&0\\ 0&0&0&0\end{array}\right)}$};
\end{tikzpicture}
\caption{\footnotesize The structure of the (bosonic) $\SU(2)^{\times4}$ subgroup of the stabilizer super group $\SU(2|2){\times}\SU(2|2)$ of the vacuum solution. The matrices above indicate the defining 8-dimensional representation of $\PSU(2,2|4)$ in $2\times2$ block form.}
\label{f1}
}

Let us work at the algebra level, initially in the undeformed theory $q=1$, and focus on one of the two copies of $\msu(2|2)$.\footnote{We will not need to worry here about issues involving which real form is relevant and so we could have equally well used the notation $\msl(2|2)$.} It was shown in \cite{Beisert:2008tw} that the simpler $\mpsu(2|2)$ super algebra admits three distinct central extensions $\mpsu(2|2)\ltimes\mathbb R^3$ . The three central extensions are determined physically by the energy, momentum and abelian charge of the states and are common to the two $\mpsu(2|2)$.

This algebra has two series of {\it short\/}  representations of dimension $4a$, $a=1,2,\ldots$, denoted $\langle a-1,0\rangle$ and $\langle 0,a-1\rangle$  \cite{Beisert:2008tw} (see also \cite{Hoare:2011nd}). These representations have 
the following decomposition under the $\msu(2)\oplus\msu(2)\subset\mpsu(2|2)$ bosonic subalgebra:
\EQ{
\langle a -1,0\rangle&=(a ,0)\oplus(a -1,1)\oplus(a -2,0)\ ,\\
\langle 0,a -1\rangle&=(0,a )\oplus(1,a -1)\oplus(0,a -2)\ ,
}
where the numbers in round brackets indicate twice the $\msu(2)$ spin. 

These representations describe bound states of giant magnons in the AdS$_5{\times}S^5$ string theory. One puts together a copy of $\langle a -1,0\rangle$ for each $\msu(2|2)$ factor giving bound states of dimension $4a \cdot4a =16a ^2$. In particular, the giant magnons are associated to these bound states where the highest spin $\frac12a $ is for the two $\SU(2)$'s that lie in the $\SU(4)$, i.e.~$\SU(2)_3$ and $\SU(2)_4$ in fig.~\ref{f1}, associated to the $S^5$.

Note that the three central charges $(C,P,K)$ are common to both centrally extended super algebras $\mpsu(2|2)$  and satisfy the shortening condition
\EQ{
C^2-PK={(a/2)}^2\ ,
\label{ms5}
}
where
\EQ{
C={\cal E}/2\ ,\qquad P=g\big(1-e^{i{\cal P}/2g}\big)\ ,\qquad K=g(1-e^{-i{\cal P}/2g}\big)\ .
}
Here, ${\cal E}$ and ${\cal P}$ are the energy and momentum, and the shortening condition~\eqref{ms5}  is just the dispersion relation  \eqref{yr2}  of the (dyonic) magnons with charge ${\cal Q}=a$.

Now we turn on the deformation $q=\exp(i\pi/k)$. 
Then, the super algebra is deformed into the quantum group and the shortening  condition becomes
\EQ{
[C]_q^2-PK=[a/2]_q^2\, ,
\label{sht}
} 
where $[x]_q=(q^x-q^{-x})/(q-q^{-1})$.

The S-matrices of the world sheet excitations and its $q$ deformation are usually presented in terms of pairs of abstract kinematic variables $x^\pm$ that satisfy the dispersion relation\footnote{The relation between the parameters used in this paper and those in \cite{Hoare:2011wr,Hoare:2013hhh} are as follows: the parameter $\xi$ in \cite{Hoare:2011wr,Hoare:2013hhh} will be denoted by $\tilde{\xi}$ here. The coupling $g$ in \cite{Hoare:2011wr,Hoare:2013hhh}, call it $\tilde g$, is related to the one used in this paper via $(1+\frac k{2\pi g})^2=1+(2\tilde g\sin(\pi/k))^{-2}$. Note that $g\to \tilde g$ as $k\to\infty$.}

\EQ{
q^{-a}\Big(x^++\frac1{x^+}+
\tilde{\xi}+ \frac{1}{\tilde{\xi}}\Big)=q^{a}\Big(x^-+\frac1{x^-}+\tilde{\xi}+ \frac{1}{\tilde{\xi}}\Big)\,,\qquad \tilde{\xi}=\frac{\lambda^{-1}-\lambda}{\lambda^{-1}+\lambda}\ .
\label{p11}
}
The dispersion relation was originally written in this form in \cite{Hoare:2011wr}, and we have expressed it in terms of $\lambda$ to anticipate the relation to the lambda model. The variables $x^\pm$ label the states in the representation specified by $a$, so that $a=1$ corresponds to the fundamental particles and $a>1$ to their bound states.
They encode the energy and momentum of a given state. To make this concrete, it is customary to define the two quantities
\EQ{
&
U^2= q^{-a}\, \frac{x^+ +\tilde{\xi}}{x^-+ \tilde{\xi}}=q^{a}\, \frac{1/x^- +\tilde{\xi}}{1/x^+ + \tilde{\xi}}\,,\qquad
V^2= q^{-a}\, \frac{\tilde{\xi} x^+ +1}{\tilde{\xi} x^- +1}= q^{a}\,\frac{\tilde{\xi}/x^-+1}{\tilde{\xi}/x^++1}\,,
\label{UandV0}
}
where the equalities follow from the dispersion relation~\eqref{p11}. 
These quantities determine the three central charges $(C,P,K)$ of the underlying symmetry algebra by means of
\EQ{
\hspace{-0.1cm}
q^{2C}=V^2\,,\quad P=\frac{i}{2}\cdot \frac{\lambda^{-1}-\lambda}{q-q^{-1}}\cdot \left(1-U^2 V^2\right)\,,\quad K=\frac{i}{2}\cdot \frac{\lambda^{-1}-\lambda}{q-q^{-1}}\cdot \left(V^{-2}-U^{-2}\right)\,,
}
which satisfy the shortening condition~\eqref{sht}.
In~\cite{Hoare:2012fc}, the relation of $U$ and $V$ with the energy and momentum was taken to be
\EQ{
U^2 =\exp\left[\frac{2\pi i{\cal P}}{k(\lambda^{-1}-\lambda)} \right]\,,\qquad
V^2 =\exp\left[\frac{2\pi i{\cal E}}{k(\lambda^{-1}+\lambda)} \right]\,,
\label{UandV2}
}
so that the shortening condition~\eqref{sht} becomes the mass shell condition~\eqref{dis} for the lambda model with ${\cal Q}=a$.
Since the eqs.~\eqref{UandV0} can be equivalently written as
\EQ{
&x^+= \tilde\xi\; \frac{1-q^a U^2}{q^a V^{-2} -1} = \frac{1}{\tilde\xi}\; 
\frac{q^a V^2-1}{1- q^a U^{-2}}\,,\qquad x^-=\tilde\xi\; \frac{U^{-2} - q^a}{q^a-V^2}= \frac{1}{\tilde\xi}\; 
\frac{q^a-V^{-2}}{U^2-q^a}\,,
}
eqs.~\eqref{UandV2} provide a mapping $x^\pm=x^\pm({\cal E},{\cal P})$. 
Notice that the quantities ${\cal E}$ and ${\cal P}$ take real values provided that we impose the reality condition $(x^+)^\ast=x^-$.

In order to define a one-to-one map between ${\cal E}$ and ${\cal P}$ and the central charges $(C,P,K)$, we restrict the arguments of $U^2$ and $V^2$ to lie in the range $(-\pi,\pi)$. Then, for each
\EQ{
|{\cal P}|< \frac{k}{2}(\lambda^{-1}-\lambda)\,,
}
the dispersion relation gives rise to two values $\pm {\cal E}({\cal P})$.
An important observation is that there are two distinct branches distinguished by the value of the momentum:
\EQ{
&
\text{magnon branch:}\quad |{\cal P}| >\frac{1}2(\lambda^{-1}-\lambda)\,a\,,\\[5pt]
&
\text{soliton branch:}\quad\;\; |{\cal P}| <\frac{1}2(\lambda^{-1}-\lambda)\,a\,.
\label{branches}
}
The two branches touch at the special values
\EQ{
|{\cal P}| =\frac{1}2(\lambda^{-1}-\lambda)\,a
\;\; \Rightarrow \;\; {\cal E} =\frac{1}2(\lambda^{-1}+\lambda)\, a\,.
\label{Special}
}
For $a>1$, this suggests the interpretation that the $a$-bound state is at threshold for decay into $a$~constituents.

In~\cite{Hoare:2012fc}, it was assumed that 
physical states correspond to the positive values of ${\cal E}$. 
However, it was already noticed in~\cite{Hoare:2012fc} that this identification of physical solutions leads to a puzzle about the nature of the poles of the S-matrix.
To be specific, let us consider the case 
of the bound states with $a=2$. Potential bound state poles of the S-matrix elements for the scattering of $a_1=a_2=1$ (fundamental particles) are found to be at $x_1^+=x_2^-$ or $x_1^-=x_2^+$. However, in order to be bona-fide bound state poles, the wave function of the bound state must be normalizable, and the normalizability condition determines the {\it physical region\/} for the kinematic variables. In \cite{Hoare:2012fc}, the physical pole for the $a=2$ bound state in the magnon branch was determined to be $x_1^+=x_2^-$. This pole corresponds to a bound state in the representation $\langle 1,0\rangle$. This state is charged under $\SU(2)_3$ and $\SU(2)_4$ which, referring to fig.~\ref{f1}, are associated to the $S^5$ component of $\text{AdS}^5\times S^5$. 
In contrast, for the soliton branch the physical bound state pole was identified as $x_1^-=x_2^+$, so that the bound state transforms in the representation $\langle0,1\rangle$. This state is charged under $\SU(2)_1$ and $\SU(2)_2$, which are associated to the $\text{AdS}^5$ component. The puzzle was that, in the classical limit, magnon/soliton solutions can only be non-trivial in the $S^5$ part of the geometry. If one tries to define a soliton in the $\text{AdS}_5$ part, the solution is singular. Therefore, the soliton bound states in the soliton branch seem to be in the wrong representation: $\langle 0,1\rangle$ rather than $\langle1,0\rangle$.

We propose to solve this puzzle simply by changing the identification of the energy and momentum in the soliton branch. Namely,\footnote{Recall that the identification of physical bound state poles in~\cite{Hoare:2012fc} involves an analytic continuation of the energy and momentum of the constituent particles.}
\EQ{
\big({\cal E}^{\text{(phys)}},{\cal P}^{\text{(phys)}}\big)= \begin{cases}
\big({\cal E},{\cal P}\big) \,,&  \text{magnon branch}\\[5pt]
\big(-{\cal E}^\ast,-{\cal P}^\ast\big) \,,& \text{soliton branch}\,,
\end{cases}
}
so that the physical states correspond to the (real) positive values of ${\cal E}^{\text{(phys)}}$. This is equivalent to using a different mapping between the kinematic variables and the energy and momentum in the magnon and soliton branches
\EQ{
x^\pm\big({\cal E}^{\text{(phys)}},{\cal P}^{\text{(phys)}}\big) = \begin{cases}
x^\pm({\cal E},{\cal P})\,,&  \text{magnon branch}\\[5pt]
1/\big[x^\mp({\cal E},{\cal P})\big]^\ast \,,& \text{soliton branch}\,.
\end{cases}
\label{mapping}
}
Repeating the analysis of \cite{Hoare:2012fc} in terms of ${\cal E}^{\text{(phys)}}$ and ${\cal P}^{\text{(phys)}}$ it turns out that the physical pole for the $a=2$ bound state is $x_1^+=x_2^-$ in both branches, so that it is always associated to the $S^5$ component of $\text{AdS}^5\times S^5$.

\vspace{0.5cm}
\FIG{
    \begin{center}
    \begin{tikzpicture}[xscale=0.4,yscale=0.45]
\draw[very thick] (-12,0) -- (12,0) -- (12,12) -- (-12,12) -- (-12,0);
\draw[thin] (-12,-0.2) -- (-12,-0.8);
\draw[thin] (12,-0.2) -- (12,-0.8);
\node at (4.5,-1.5) (a2) {\footnotesize$\frac12(\lambda^{-1}-\lambda)a$};
\node at (-4.5,-1.5) (a5) {\footnotesize$-\frac12(\lambda^{-1}-\lambda)a$};
\node at (12,-1.5) (a1) {\footnotesize$\frac12(\lambda^{-1}-\lambda)k$};
\node at (-12,-1.5) (a1) {\footnotesize$-\frac12(\lambda^{-1}-\lambda)k$};
\node at (0,-3.5) (a1) {${\cal P}^{\text{(phys)}}$};
\node at (-14.2,7) (a1) {${\cal E}^{\text{(phys)}}$};
\draw[thin] (4,-0.8) -- (4,4.515);
\draw[thin] (-4,-0.8) -- (-4,4.515);
\draw[thin] (-12.8,4.515) -- (4,4.515);
\filldraw[black] (4,4.515) circle (4pt);
\filldraw[black] (-4,4.515) circle (4pt);
\node at (-15.8,4.515) (a3) {\footnotesize$\frac12(\lambda^{-1}+\lambda)a$};
\node at (0,6) (b1) {\footnotesize marginally stable};
\draw[thin,->] (b1) -- (3.5,4.7);
\draw[thin,->] (b1) -- (-3.5,4.7);
\draw[very thick,red] plot[smooth] coordinates {(-4.,4.51514)   (-3., 3.65017) (-2., 2.86478)   (-1., 2.2606)   (0., 2.01838)  (1., 2.2606)  (2., 2.86478)  (3., 3.65017)  (4., 
  4.51514)};
\draw[very thick,blue] plot[smooth] coordinates   {(4., 4.51514)  (4.57143, 5.02511)  (5.14286, 5.53909)  (5.71429, 
  6.05263)  (6.28571, 6.56165)  (6.85714, 7.06206)  (7.42857, 
  7.54933)  (8., 8.01805)  (8.57143, 8.46154)  (9.14286, 
  8.87133)  (9.71429, 9.23672)  (10.2857, 9.54471)  (10.8571, 
  9.78052)  (11.4286, 9.92956)  (12., 9.98065)} ;
\draw[very thick,blue] plot[smooth] coordinates   {(-4., 4.51514)  (-4.57143, 5.02511)  (-5.14286, 5.53909)  (-5.71429, 
  6.05263)  (-6.28571, 6.56165)  (-6.85714, 7.06206)  (-7.42857, 
  7.54933)  (-8., 8.01805)  (-8.57143, 8.46154)  (-9.14286, 
  8.87133)  (-9.71429, 9.23672)  (-10.2857, 9.54471)  (-10.8571, 
  9.78052)  (-11.4286, 9.92956)  (-12., 9.98065)} ; 
  \node[blue] at (5,9) (b1) {$|x^+|>1$};
   \node[blue] at (-5,9) (b1) {$|x^+|>1$};  
   \node[red] at (0,1) (b2) {$|x^+|<1$};
   \end{tikzpicture}
    \end{center}
\caption{\footnotesize The solution of the dispersion relation for the bound state labeled by~$a$ gives the energy as a function of the momentum. The momentum is only valued in the finite region as indicated. The soliton (red) and magnon branches (blue) are shown. In the sigma model limit, $\lambda\to1$ ($k\to\infty$) and the soliton branch disappears. In the opposite limit $\lambda\to0$, the magnon branch disappears. The magnon and soliton branches are on disconnected sheets of the S-matrix rapidity torus. However, the energy and momentum  ``touch'' at points of marginal stability.
\label{f2}
}}

In the magnon branch, the resulting physical solutions have $|x^+|>1$ which, using the conventions of~\cite{Hoare:2011wr,Hoare:2012fc}, means that that they are in the sheet ${\cal R}_0$. In contrast, in the soliton branch they have $|x^+|<1$ and, thus, they lie on ${\cal R}_{\pm2}$. In both cases,  $\IM(x^+)>0$.\footnote{It is important to point out that mapping between the underlying S-matrix parameters $x^\pm$ and the energy and momentum does not affect the S-matrix axioms: unitarity, crossing and the Yang-Baxter equation.} Notice that our identification of physical solutions puts the magnon and soliton branches on disconnected sheets of the S-matrix rapidity torus. In fact,
${\cal R}_0$ and ${\cal R}_{\pm2}$ are related by means of the antipode map $x^\pm\to 1/x^\pm$ of the quantum group.
In particular, this means that the two branches do not actually touch each other. In fact, the special values~\eqref{Special} correspond to the following asymptotic values of the kinematical parameters:
\EQ{
&
q^{-a}\, x^+ \to -\infty
 \;\; \Rightarrow\;\; \frac{2\, {\cal E}}{\lambda^{-1}+\lambda}=\frac{2\, {\cal P}}{\lambda^{-1}-\lambda}=a\,,\\[5pt]
&
q^{-a}\, (x^+ +1/\tilde{\xi}) \to 0 \;\; \Rightarrow\;\; \frac{2\, {\cal E}}{\lambda^{-1}+\lambda}=-\frac{2\, {\cal P}}{\lambda^{-1}-\lambda}=a}
which, since $|\widehat{\xi}|<1$, are in the magnon branch ($|x^+|>1$), and
\EQ{
&
q^{a}\, x^+ \to 0
 \;\; \Rightarrow\;\; \frac{2\, {\cal E}}{\lambda^{-1}+\lambda}=\frac{2\, {\cal P}}{\lambda^{-1}-\lambda}=-a\\[5pt]
&
q^{-a}(x^+ +\widetilde{\xi}) \to 0 \;\; \Rightarrow\;\; \frac{2\, {\cal E}}{\lambda^{-1}+\lambda}=-\frac{2\, {\cal P}}{\lambda^{-1}-\lambda}=-a\,,
}
which are in the soliton branch ($|x^+|<1$).

In section~\ref{s3} we shall show that the spectrum of physical solutions agrees with the soliton solutions (giant magnons) of the lambda model. The latter will be specified by their energy and momentum, and an additional charge ${\cal Q}$ quantized so that it equals~$a$, the positive integer that labels the representation $\langle a-1,0\rangle$. 
The resulting picture is the following. As shown in fig.~\ref{f2}, in the sigma model limit $\lambda\to 1$ ($q=1$) the soliton branch disappears and $a=1,2,\ldots,\infty$: there is only a ``magnon branch" where all the states have $|x^+|>1$ (${\cal R}_0$). In the deformed theory  this tower of states becomes truncated $a=1,2,\ldots,k-1$ and splits in two branches: magnon and soliton, the latter with $|x^+|<1$ (${\cal R}_{\pm2}$). In the opposite (sine-Gordon) limit $\lambda\to0$ the magnon branch disappears.
What is unusual, is that which of the two branches is relevant depends on the momentum of the state  \cite{Hoare:2012fc,Hoare:2013hhh}. This state of affairs could not be realized in a relativistic theory where a state with non vanishing momentum is just a boost of the same state at rest.

\section{Gauge fixing and the Hofman-Maldacena limit}\label{s2}

In this section, we describe how the world sheet theory can be gauge fixed. The approach we take is to generalize the conformal gauge fixing approach of Hofman and Maldacena \cite{Hofman:2006xt}  of the $\text{AdS}_ 5{\times} S^5$ world sheet sigma model (related to the Pohlmeyer reduction of Tseytlin and Grigoriev \cite{Grigoriev:2007bu}). This starts by fixing the world sheet metric $g_{\mu\nu}=e^\phi\eta_{\mu\nu}$ and, in this approach, the Virasoro constraints are imposed by hand. 

For the sigma model, the gauge fixing relies on the existence of isometries in the spacetime, in particular, shifts  $t\to t+c$ and rotations  $\phi\to\phi-c$, where $t$ is the usual time of global AdS$_5$ and $\phi$ is the angular coordinate on the equator of $S^5$. These isometries allow one to identify the physical configuration space with 
transverse excitations (to be made precise) around the BMN solution \cite{Berenstein:2003gb} for which the sigma model field is
\EQ{
f_0=\exp[2\tau\Lambda]\ .
\label{BMN}
}
This solution describes a point like string moving along a null geodesic corresponding to motion along the equator of the $S^5$. The null geodesic is described algebraically by a constant bosonic element of the Lie super algebra $\mpsu(2,2|4)$. This element has grade 2 under the $\mathbb Z_4$ automorphism and, up to conjugation can be chosen to be
\EQ{
\Lambda&=\mu(\Lambda_1-\Lambda_2)\ ,\\[5pt] \Lambda_1&=\frac{i}2{\footnotesize \left(\begin{array}{cc|cc} I_2 &\  0\  & \ 0\ & \ 0\ \\ \ 0\  & -I_2 & \ 0\ & \ 0\ \\ \hline \ 0\ &\ 0\ &\ 0\ &\ 0\  \\ \ 0\ &\ 0\ &\ 0\ &\ 0\ \end{array}\right)}\ ,\qquad
\Lambda_2=\frac{i}2{\footnotesize \left(\begin{array}{cc|cc}  \  0\  & \ 0\phantom{a}  & \ 0\ & \ 0\ \\ \ 0\  & \ 0 \phantom{a} & \ 0\ & \ 0\ \\ \hline \ 0\ &\ 0\phantom{a} &I_2&0\\ \ 0\ &\ 0\phantom{a} &\ 0\ &-I_2\end{array}\right)}\ .
\label{vcc}
}
Here, $\mu$ is a constant with unit mass dimension. The Lie super algebra element $\Lambda$ gives rise to the orthogonal decomposition
\EQ{
{\mathfrak  f}=\text{Ker}\big[\text{ad}(\Lambda)\big]\oplus
  \text{Im}\big[\text{ad}(\Lambda)\big]\equiv {\mathfrak
  f}^\perp\oplus{\mathfrak f}^\parallel\ .
\label{dec}
}
The component $\Lambda_1$ here describes isometries corresponding to the usual time coordinate in global AdS$_5$ while $\Lambda_2$ describes rotation of the equator of the $S^5$. The fact that the geodesic associated to $\Lambda$ is 
null corresponds to 
\EQ{
\STr(\Lambda\Lambda)=0\ .
}
This condition also ensures that the Virasoro constraints~\eqref{vc} are satisfied by the BMN solution. 

The physical gauge-fixed configuration space can be related in a nice way to the Lax equations of the model and its linear system. The BMN solution~\eqref{BMN} corresponds to the ``vacuum" solution of the linear system
\EQ{
\Psi_0(x;z)=\exp\big[-(z^2\sigma^++z^{-2}\sigma^-)\Lambda\big]\ ,
\label{vlsy}
}
via~\eqref{eyy2}, where $\sigma^\pm=\tau\pm\sigma$ are light cone coordinates, so that
$\AA_{0\,\pm}=-\partial_\pm \Psi_0 \Psi_0^{-1}= z^{\pm2}\Lambda$. The gauge fixed configuration space will be identified as the orbit of {\it dressing transformations\/} acting on this vacuum solution. 

Before we discuss the dressing transformations, let us complete the discussion of the (undeformed) sigma model by considering the Hamiltonian.
The Virasoro constraints seem to imply that the world sheet Hamiltonian vanishes. The puzzle then is what generates time translations in the gauge fixed theory? The resolution is that the gauge fixing procedure is explicitly time dependent, involving as it does the BMN solution, and this leads to a shift in the na\"\i ve vanishing Hamiltonian by the corresponding Noether charge that generates the BMN solution \cite{Evans:1993dq}. If we denote the charge generating the translations $t\to t+c$ as $\Delta$ and rotations $\phi\to\phi+c$ as $J$, this identifies the world sheet energy as ${\cal E}=\Delta-J$. Note that $\Delta$ is a space time energy and $J$ a global charge corresponding to one of the isometries of the $S^5$. Writing the $F/G$ sigma model in terms of the group valued field $f\in F$ and a gauge field $B_\mu\in\mathfrak g$, we have
\EQ{
\Delta&=2\mu g\int_{-\pi}^\pi d\sigma\,\STr\big[\Lambda_1 (\partial_0ff^{-1}+fB_0f^{-1})\big]\ ,\\
J&=2\mu g\int_{-\pi}^\pi d\sigma\,\STr\big[\Lambda_2 (\partial_0ff^{-1}+fB_0f^{-1}\big]\ .
\label{scs}
}
In the dual ${\cal N}=4$ theory, $\Delta$ is identified with the scaling dimension and $J$ with one of the $R$ charges.

Now we turn to the lambda model. The gauge fixing procedure is exactly the same as we have already described for the sigma model. One fixes conformal gauge and then solves the Virasoro constraints by hand by taking the same solution of the linear solution \eqref{vlsy}. 
Using~\eqref{eyy}, this gives the ``vacuum" solution 
\EQ{
\CF_0=\exp\big[2(\lambda^{-1}-\lambda)\sigma\Lambda\big]\ .
\label{vlm}
}
It follows that in the lambda model the interpretation of the vacuum solution is completely different. It is a static closed string that wraps around a cycle on the lambda background. Closed string boundary conditions $\CF(-\pi)=\CF(\pi)$ require the quantization condition
\EQ{
\mu(\lambda^{-1}-\lambda)\in\mathbb Z\ .
}
The fact that a momentum mode of the string in AdS$_5{\times}S^5$ becomes a winding mode in the lambda model is characteristic of a T duality.

In the sigma model case, the gauge fixing procedure relied on the existence of isometries in the background and associated Noether charges on the world sheet. The background spacetime of the lambda model is complicated and the existence of symmetries is difficult to see explicitly. We show in appendix \ref{a2} that the lambda model does have Noether symmetries whose charges play the same role as $\Delta$ and $J$ in the gauge fixing procedure and which reduce to those in the sigma model limit. It is an interesting question as to whether there is a target space interpretation of these world sheet symmetries given that the lambda background appears not to have any isometries. The question of how Noether symmetries of the world sheet can be pushed up to the space time certainly deserves further investigation.\footnote{We thank the referee of this paper for raising this point.}

It is, perhaps, not surprising that such charges exist, after all the world sheet theory is integrable and there are many charges, some related to local and some to non-local conserved currents. In the theory of integrable systems an infinite set of both local and non-local conserved currents can be constructed by the process of {\it abelianizing\/} the Lax connection around one of its poles. In the present case this is either at $z=0$ or $\infty$. The details are in appendix \ref{a1}. In particular, the charges corresponding to the local conserved currents that we need can be extracted from the ``right monodromy"
\EQ{
{\cal W}(z)=\Psi^{-1}(\sigma=-\pi;z)\Psi(\sigma=\pi;z)\ ,
\label{mon}
}
as
\EQ{
\mathfrak Q(z)=\STr\big(\Lambda\log {\cal W}(z)\big)\ .
}
We can think of $\mathfrak Q(z)$ as being a generating function for an infinite set of charges.
Specifically the physical energy and momentum are equal to
\EQ{
{\cal E}&=\frac {k}{4\pi}(\lambda^{-1}+\lambda)\STr\big[\Lambda \log{\cal W}(\lambda^{-1/2}){\cal W}^{-1}(\lambda^{1/2})\big]\ ,\\
{\cal P}&=\frac{k}{4\pi}(\lambda^{-1}-\lambda)\STr\big[\Lambda \log{\cal W}(\lambda^{-1/2}){\cal W}(\lambda^{1/2})\big]\ .
\label{iuu}
}

In the sigma model limit, $k\to\infty$ with $g$ fixed, that is $\lambda\to1$, the energy becomes identified with the Noether charge of the sigma model $\Delta-J$ defined in \eqref{scs}. In addition, the momentum ${\cal P}$ has the interpretation of a winding number for the group field $f(\sigma)$:
\EQ{
{\cal P}\longrightarrow 2g\STr\big[\Lambda(\log f(-\pi)-\log f(\pi))\big]\ .
}
Of course in the string theory context, the group field should be periodic and this means that the overall momentum should vanish. On the contrary, in the lambda model it is the energy which has the interpretation as winding:
\EQ{
{\cal E}=\frac k{4\pi}(\lambda^{-1}+\lambda)\STr\big[\Lambda(\log\CF(-\pi)-\log\CF(\pi)\big]\ .
\label{energy}
}

\subsection{The gauge fixed configuration space}

Since they share the same linear system, the gauge fixed configuration space of the sigma and lambda models will be identified with a special set of transformations, the dressing transformations, which act on the vacuum solution \eqref{vlsy} in such a way as to preserve the Virasoro constraints.

A dressing transformation is associated to an element of the loop group $g(z)$ for which it is assumed 
that there is a factorization of the form
\EQ{
g(z)=g_-(z)^{-1}g_+(z)\ ,
}
where $g_+(z)$ and $g_-(z)$ are formal series in $z$ and $z^{-1}$, respectively. We shall choose the normalization condition $g_-(\infty)=1$  (for a review, see the book \cite{BBT}). Notice that this factorization is unique only if $g(z)$ is close enough to the identity, and we require that $g_+(z)$ and $g_-(z)$ are invertible and analytic around $z=0$ and $z=\infty$, respectively, so that it can be understood in terms of a Riemann-Hilbert problem. However, any possible factorization gives rise to a dressing transformation. In particular, it is remarkable that
the construction of soliton solutions involve non trivial factorizations of $g(z)=1$ ``with zeros", where $g_+(z)$ and $g_-(z)$ exhibit simple poles~\cite{BBT}.

Let $\Psi$ be a solution of the linear system and define
\EQ{
\Theta(x;z)=\Psi(x;z)g(z)\Psi(x;z)^{-1}\ .
}
At each point in spacetime $x$, one then performs a factorization as above
\EQ{
\Theta(x;z)=\Theta_-(x;z)^{-1}\Theta_+(x;z)\ .
\label{hee}
}
Then it follows that
\EQ{
\Psi^g(x;z)=\Theta_\pm(x;z)\Psi(x;z)g_\pm(z)^{-1}
\label{jff}
}
also satisfies the linear system for either choice of sign.

The physical gauge fixed phase space is then identified with the orbit of the dressing group acting on the vacuum solution $\Psi_0$ given by~\eqref{vlsy}. The orbit can be parameterized by a set of fields: $\gamma$, a group element of $G\subset F$, and $\psi_\pm$, two Grassmann fields in $\mf^{(1)}$ and $\mf^{(3)}$ that lie in the image of $\text{ad}(\Lambda)$. Along this orbit, the Lax connection takes the form
\EQ{
\AA_+=\gamma^{-1}\partial_+\gamma+z\psi_++z^2\Lambda\ ,\qquad
\AA_-=z^{-1}\gamma^{-1}\psi_-\gamma +z^{-2}\gamma^{-1}\Lambda\gamma\ .
\label{gfLax}
}
The equations of motion of these fields are the non abelian Toda equations.
These are also the Lax equations of the gauge fixed Pohlmeyer/sine Gordon theory \cite{Grigoriev:2007bu,Mikhailov:2007xr,Hollowood:2011fq,susy-flows}.
Our gauge fixing prescription is equivalent to imposing the constraints
\EQ{
&J_+^{(2)}= \Lambda\,,\qquad J_-^{(2)}= \gamma^{-1}\Lambda\gamma\,,\qquad J_+^{(0)}=\gamma^{-1}\partial_+ \gamma\,,\qquad J_-^{(0)}=0\,,\\[5pt]
&J_+^{(1)}= \psi_+ \,,\qquad J_-^{(1)}= 0\,,\qquad
J_+^{(3)}= 0\,,\qquad J_-^{(3)}= \gamma^{-1}\psi_- \gamma \,,
\label{gfcond}
}
used by Grigoriev and Tseytlin in the context of the Pohlmeyer reduction of the sigma model~\cite{Grigoriev:2007bu}.
Once the Lax connection takes the form~\eqref{gfLax}, the connection with the dressing transformations was explicitly worked out in~\cite{Hollowood:2011fq}.

It is important to stress that the gauge fixing procedure fixes all the gauge symmetries including kappa symmetry. The only residual symmetries are the (bosonic) global gauge transformations $\AA_\pm\to U\AA_\pm U^{-1}$, where $U\in G$ and $U\Lambda U^{-1}= \Lambda$.

\subsection{The Hofman-Maldacena limit}\label{s2.1}

A particularly interesting limit is the Hofman-Maldacena (HM) limit \cite{Hofman:2006xt}, which in the sigma model involves focusing on states with very large charges 
$\Delta,J\to\infty$ but with $\Delta-J$ fixed. So in the world sheet theory this means with large Noether charges $\Delta$ and $J$ but finite energy ${\cal E}=\Delta-J$. In the gauge-gravity correspondence such string states are associated to operators on the gauge theory side that are single trace operators built from a high power of a given complex scalar field---picked out by the choice of the charge $J$---and a finite number of other fields. In this limit, the operators can be put into correspondence with the states of a spin chain in the thermodynamic limit.

The HM limit \cite{Hofman:2006xt} corresponds to taking the mass scale $\mu\to\infty$. 
It is very convenient to then absorb this scale into the spacetime coordinates on the world sheet and define new re-scaled coordinates $(t,x)=\mu(\tau,\sigma)$ and set $\mu=1$ in the definition of $\Lambda$ in \eqref{vcc}. The original spatial coordinate $\sigma\in[-\pi,\pi]$ was periodic while the new re-scaled coordinate $x\in[-\infty,\infty]$. Effectively, the world sheet decompactifies. 

In the lambda theory, the HM limit of the vacuum configuration describes a string that wraps an infinite number of times around the lambda background, as is clear from \eqref{vlm}. 

\section{Giant magnons}\label{s3}

In this section, we will consider the soliton solutions on the world sheet known as giant (dyonic) magnons. We should emphasize that since the sigma and lambda models have the same linear system, the giant magnons are common to both.
These solutions can be efficiently constructed from the linear system via the dressing method \cite{Kalousios:2006xy,Spradlin:2006wk,Hollowood:2009tw,Hollowood:2009sc,Hollowood:2011fq,Hollowood:2010dt,Hollowood:2011fm}.\footnote{The dressing method originally goes back to \cite{Zakharov:1973pp}.} For us, this procedure has the added advantage that it yields the solutions in both the sigma and lambda models in one go. In fact it also yields the solution in the associated Pohlmeyer/sine-Gordon theory which describes the world sheet theory in sine-Gordon limit ($g\to\infty$, $\lambda\to0$ with $k$ fixed). This is the limit where the gauge fixed theory becomes relativistic.  

The magnon/soliton solutions are best constructed via a specific kind of dressing transformation of the vacuum solution \eqref{vlsy}. The fact that they are dressing transformations manifests the fact that they lie in the gauge-fixed configuration space. The special kind of dressing transformations are defined as in \eqref{jff} but with $g(z)=1$. What makes them non trivial is that they are dressing transformations ``with zeros" \cite{BBT}.

Solitons in 
the AdS$_ 5{\times} S^5$ semi symmetric space are constructed in~\cite{Hollowood:2011fq} following the original approach of~\cite{Harnad:1983we}. The collective coordinates of a soliton consist of a $(4|4)$ constant vector. The first 4 components of the vector are Grassmann while the second 4 components are ordinary $c$-numbers. This means that the soliton solution for the group fields and wave function has the structural form
\EQ{
f,\CF,\Psi=\left(\begin{array}{c|c} \text{fermionic}^{2} & \text{fermionic}\\\hline \text{fermionic} & \text{bosonic} \end{array}\right)\ .
\label{vcc2}
}
So the part of the solution in the AdS part of the geometry is a bosonic quantity that is a composite---at least quadratic---of the Grassmann collective coordinates.

Since we will be interested in relating the solitons to a semi-classical limit of the quantum theory, it is sufficient for us to consider purely bosonic solutions. They lie entirely in the subgroup $\SU(4)\subset\text{PSU}(2,2|4)$ associated to the $S^5$ part of the geometry. These solitons are precisely those constructed in \cite{Hollowood:2010dt} in terms of the symmetric space $S^5=\SU(4)/\Sp(4)$. 
In the $\SU(4)$ subspace, the solitons have a collective coordinate in the form of a constant 4-vector $\Bvarpi$. Using global symmetries, we can bring $\Bvarpi$ into the form,
\EQ{
\Bvarpi=(1,0,1,0)\ .
}
Note that $\SU(2)_3$ in fig.~\ref{f1}, respectively $\SU(2)_4$, acts on the first two, last two, components of $\Bvarpi$.
The soliton also has an associated complex kinematic parameter $\xi=e^{-\theta-i\alpha}$  whose significance will emerge. 
We then define the following $\mathbb Z_2$ action on the giant magnon's data: $\{\xi_i\}=\{\xi,-\xi^*\}$ and $\{\Bvarpi_i\}=\{\Bvarpi,\tilde{\cal K}\Bvarpi^*\}$, $i=1,2$, where\footnote{This $\mathbb Z_2$ action is a subgroup of the $\mathbb Z_4$ automorphism of the semi-symmetric space. Once the fermions are set to zero only a $\mathbb Z_2\subset\mathbb Z_4$ remains.
Recall that the AdS$_ 5{\times} S^5$ solitons constructed in~\cite{Hollowood:2011fq} involve four kinematic parameters $\{\tilde\xi_i\}$ so that $\tilde\xi_1^2=\tilde\xi_3^2=\xi$ and $\tilde\xi_2^2=\tilde\xi_4^2=-\xi^\ast$.}
\EQ{
\tilde{\cal K}={\footnotesize\MAT{\,0\, & -1 & \,0\, & 0\\ 1 & 0 & 0 & 0\\ 0 & 0 & 0 & -1\\ 0 & 0 & 1 & 0}}\ .
}
Then we define $\BF_i=\Psi_0\big(\sqrt{\xi_i^*}\big)\, \Bvarpi_i$, where the latter is the vacuum solution in the $\SU(4)$ subspace:
\EQ{
\Psi_0(z)=\exp[(z^2x^++z^{-2}x^-)\Lambda_2]\ .
}
The dressing transformation associated to this data then takes the form\EQ{
\Theta(z)=1+\sum_{ij}\frac{\BF_i\Gamma_{ij}^{-1}\BF_j^\dagger}{z^2-\xi_j}\ ,
\label{ruw}
}
where
\EQ{
\Gamma_{ij}=\frac{\BF_i^*\cdot \BF_j}{\xi_i-\xi_j^*}\ .
}
For $\Bvarpi=(1,0,1,0)$, we can write the dressing transformation explicitly as
\EQ{
\Theta(x;z)={\small1+\frac{\xi-\xi^*}{e^{X+X^*}+e^{-X-X^*}}\MAT{\frac{e^{-X-X^*}}{z^2-\xi} & 0 & \frac{e^{X-X^*}}{z^2-\xi} & 0\\ 
0 & \frac{e^{-X-X^*}}{z^2+\xi^*} & 0&\frac{e^{-X+X^*}}{z^2+\xi^*} \\
\frac{e^{-X+X^*}}{z^2-\xi} & 0 & \frac{e^{X+X^*}}{z^2-\xi} & 0 \\ 
0&\frac{e^{X-X^*}}{z^2+\xi^*}&0&\frac{e^{X+X^*}}{z^2+\xi^*}}}\ ,
\label{hsv}
} 
where $X=i(\xi x^++\xi^{-1}x^-)/2$.
The block form reflects the fact that the most general solution is valued in a subgroup $\SU(2){\times}\SU(2)$ of $\SU(4)$ (that is an $S^3\subset S^5$ as one expects for the dyonic giant magnon \cite{Chen:2006gea,Dorey:2006dq}). The form of the dressing transformation for generic values of $\Bvarpi$ can be found by performing a transformation $\Theta(x;z)\to U \Theta(x;z) U^\dagger$ with $U\in \SU(2)_3\times \SU(2)_4$, which is equivalent to $\Bvarpi\to U\Bvarpi$. For the lambda model field $\CF$, this transformation is a global gauge transformation, which is the residual symmetry left by our gauge fixing conditions.

The group valued fields in the sigma and lambda models are then given as in \eqref{eyy} and \eqref{eyy2} as
\EQ{
f(x)&=\exp(2 t\Lambda)\Theta(x,1)^{-1}\ ,\\[5pt]
\CF(x)&=\Theta(x;\lambda^{1/2})\exp\big[2 x(\lambda^{-1}-\lambda)\Lambda\big]\Theta(x;\lambda^{-1/2})^{-1}\ .
}
These solutions include the vacuum component in the AdS part of the geometry.
The Pohlmeyer/sine-Gordon group valued field is also determined simply as
\EQ{
\gamma(x)=\Theta(x;0)^{-1}\ .
}
One can readily verify from these explicit forms that the parameter $\theta$ is the rapidity of the solution, determining the velocity via $v=\tanh\theta$. The parameter $\alpha$ determines the internal angular velocity of the giant magnon.

The procedure of \cite{Hollowood:2010dt} does not necessarily ensure that $\det\, f =\det\, \CF=\det\, \gamma=1$, so that they take values in $SU(4)$. Therefore, it requires a compensating scalar factor that, in our case, amounts simply to the change
\EQ{
\Theta(x;z) \to \left[\frac{z^2-\xi}{z^2-\xi^\ast}\cdot \frac{z^2+\xi^\ast}{z^2+\xi}\right]^{1/4}\, \Theta(x;z)\,,
\label{Sfactor}
}
so that $\det\,  \Theta(x;z)=1$.
This compensating factor contributes neither to the monodromy nor to the conserved charges.

\pgfdeclareimage[interpolate=true,width=5cm]{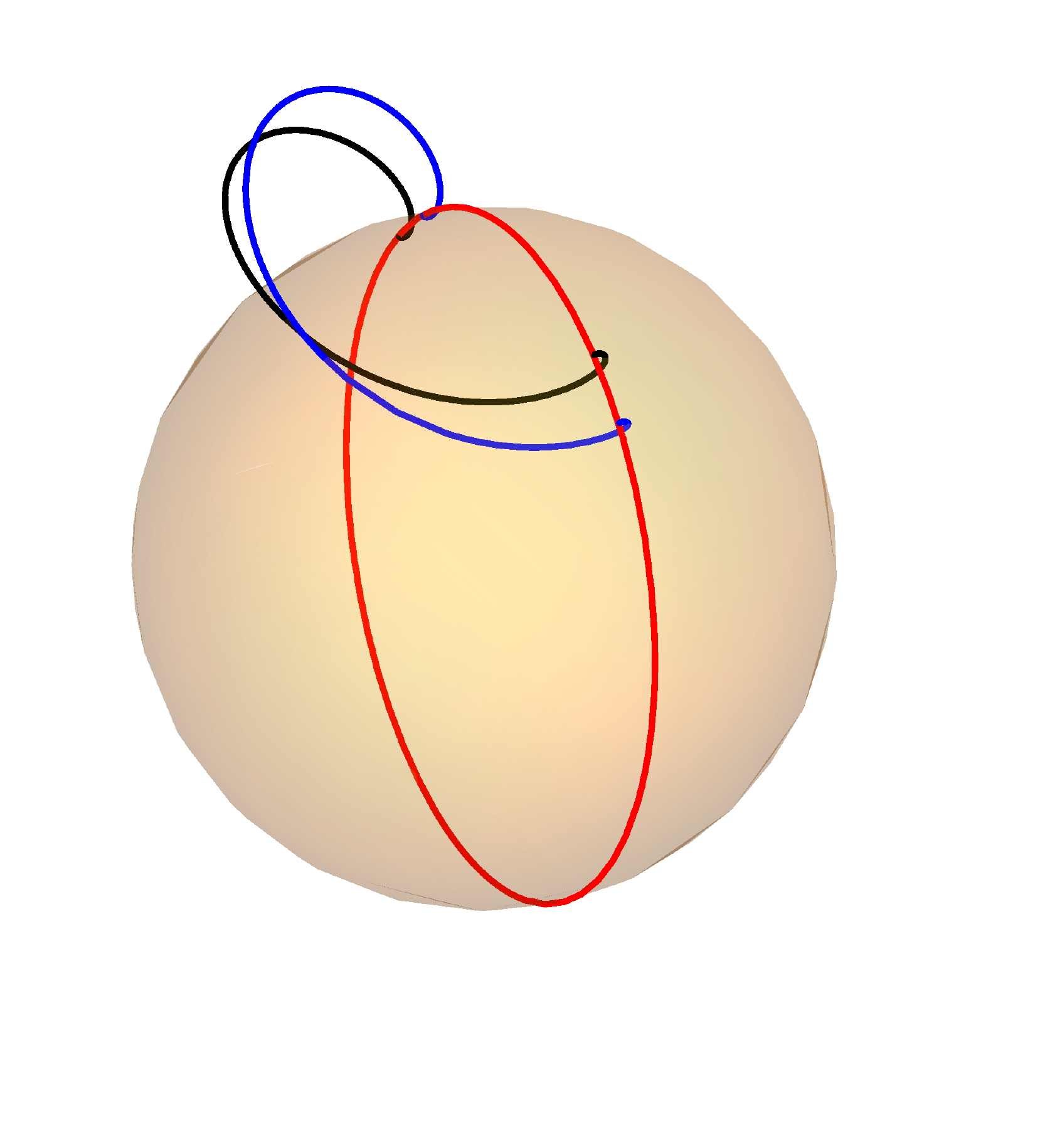}{fig1}
\pgfdeclareimage[interpolate=true,width=5cm]{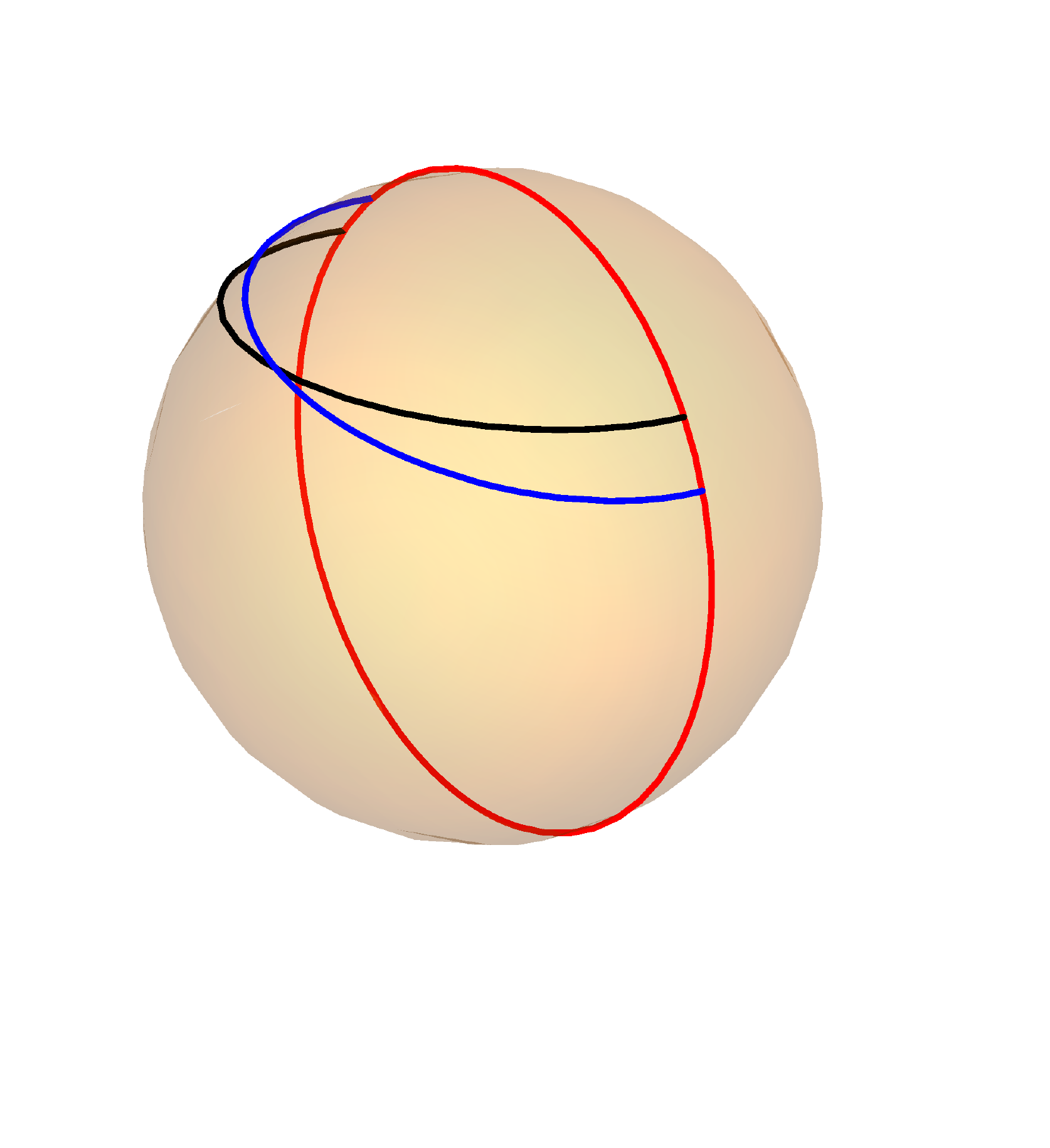}{fig2}
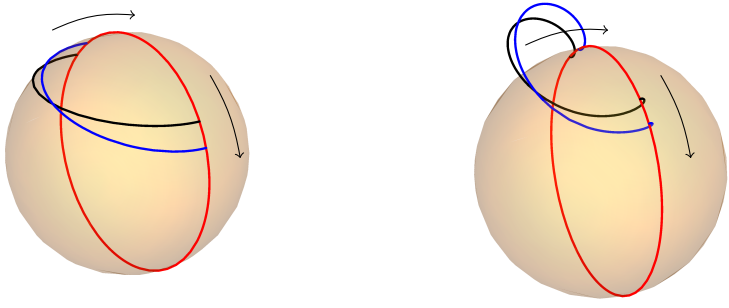
\begin{figure}
\begin{center}
\begin{tikzpicture}[scale=1]
\pgftext[at=\pgfpoint{0cm}{0cm},left,base]{\pgfuseimage{fig2}} 
\draw[->] (3.4,4.1)  to[out=-60,in=100]  (3.8,3);
\draw[->] (1.3,4.7) to[out=25,in=175] (2.4,4.9);
\end{tikzpicture}
\hspace{1cm}
\begin{tikzpicture}[scale=1]
\pgftext[at=\pgfpoint{0cm}{0cm},left,base]{\pgfuseimage{fig1}} 
\draw[->] (3.1,4.1)  to[out=-60,in=100]  (3.5,3);
\draw[->] (1.3,4.5) to[out=25,in=175] (2.4,4.7);
\end{tikzpicture}
\caption{\footnotesize Stereographic images of the giant magnon (left) and dyonic giant magnon (right) solutions at two nearby times (black and blue). The strings end on the orbit of the BMN solution shown in red. The endpoints move along this orbit at the speed of light. The giant magnon obtained by taking $\alpha=\frac\pi2$ takes values in an $S^2\subset S^5$ shown in brown.}
\label{f3}
\end{center}
\end{figure}

For the sigma model, we can extract directly the coordinates on $S^5$ by defining the gauge invariant field \cite{Arutyunov:2009ga}
\EQ{
\tilde f=f\tilde{\cal K}f^T={\footnotesize\MAT{0 & -Y_3 & -i Y_1^* & -iY_2^*\\ Y_3&0&iY_2&-iY_1 \\ iY_1^*&-iY_2&0&-Y_3^* \\ iY_2^*&iY_1&Y_3^*&0}\ .}
} 
The $Y_i$ are then complex coordinates on $S^5$, $|Y_1|^2+|Y_2|^2+|Y_3|^2=1$.
The giant magnon solution has $Y_1=0$ and so, as already noted, is valued in $S^3\subset S^5$:
\EQ{
Y_2=\frac{2i\eta(\xi-\xi^*)}{(1-\xi)(1+\xi^*)\big(e^{2X}+e^{-2X^*}\big)}\ ,\quad
Y_3=e^{-2it}\frac{\eta^{-1}e^{2X}+\eta e^{-X^*}}
{e^{2X}+e^{-2X^*}}\ ,
}
where
\EQ{
\eta=\sqrt{\frac{(1-\xi)(1+\xi^*)}{(1+\xi)(1-\xi^*)}}\ .
}
This is precisely the dyonic giant magnon solution of \cite{Chen:2006gea,Dorey:2006dq}. The ordinary giant magnon is obtained by setting the parameter $\alpha=\pi/2$. In this limit, $Y_2$ becomes real and the solution takes values in an $S^2\subset S^3\subset S^5$.
In fig.~\ref{f3} we show the ordinary magnon and its dyonic generalization as a stereographic projection of $S^3$ to $\mathbb R^3$.\footnote{Explicitly $x=\RE Y_2/(1+\IM Y_2)$, $y=\RE Y_3/(1+\IM Y_2)$ and $z=\IM Y_2/(1+\IM Y_2)$.}
The circle corresponding to the  BMN solution is shown in red. Note that the magnon solutions are open strings that end on this circle.

The resulting expression for the lambda model field $\CF$ is quite cumbersome and so we shall simply provide a picture by taking one of the $\SU(2)$ factors (so topologically an $S^3$) of an illustrative solution and stereographically plotting  it in $\mathbb R^3$ in fig.~\ref{f4}.\footnote{Recall that the lambda background is not a geometrical coset i.e.~a right coset but a left/right coset, so visualizing how the deformed giant magnon wraps this manifold is more subtle.}

The giant magnon in the Pohlmeyer/sine-Gordon takes the form
\EQ{
\gamma={\footnotesize\frac1{1+e^{2(X+X^*)}}\MAT{1+e^{-2i\alpha+2(X+X^*)} &0&(e^{-2i\alpha}-1)e^{2X^*}&0\\ 0&e^{2i\alpha}+e^{2(X+X^*)}&0&(e^{2i\alpha}-1)e^{2X^*}\\ (e^{-2i\alpha}-1)e^{2X}&0&e^{-2i\alpha}+e^{2(X+X^*)}&0\\ 0&(e^{2i\alpha}-1)e^{2X^*}&0&1+e^{2i\alpha+2(X+X^*)}}}\ .
}
This can be interpreted as a pair of complex sine-Gordon solitons in each of the $\SU(2)$ sectors.
\pgfdeclareimage[interpolate=true,width=7cm]{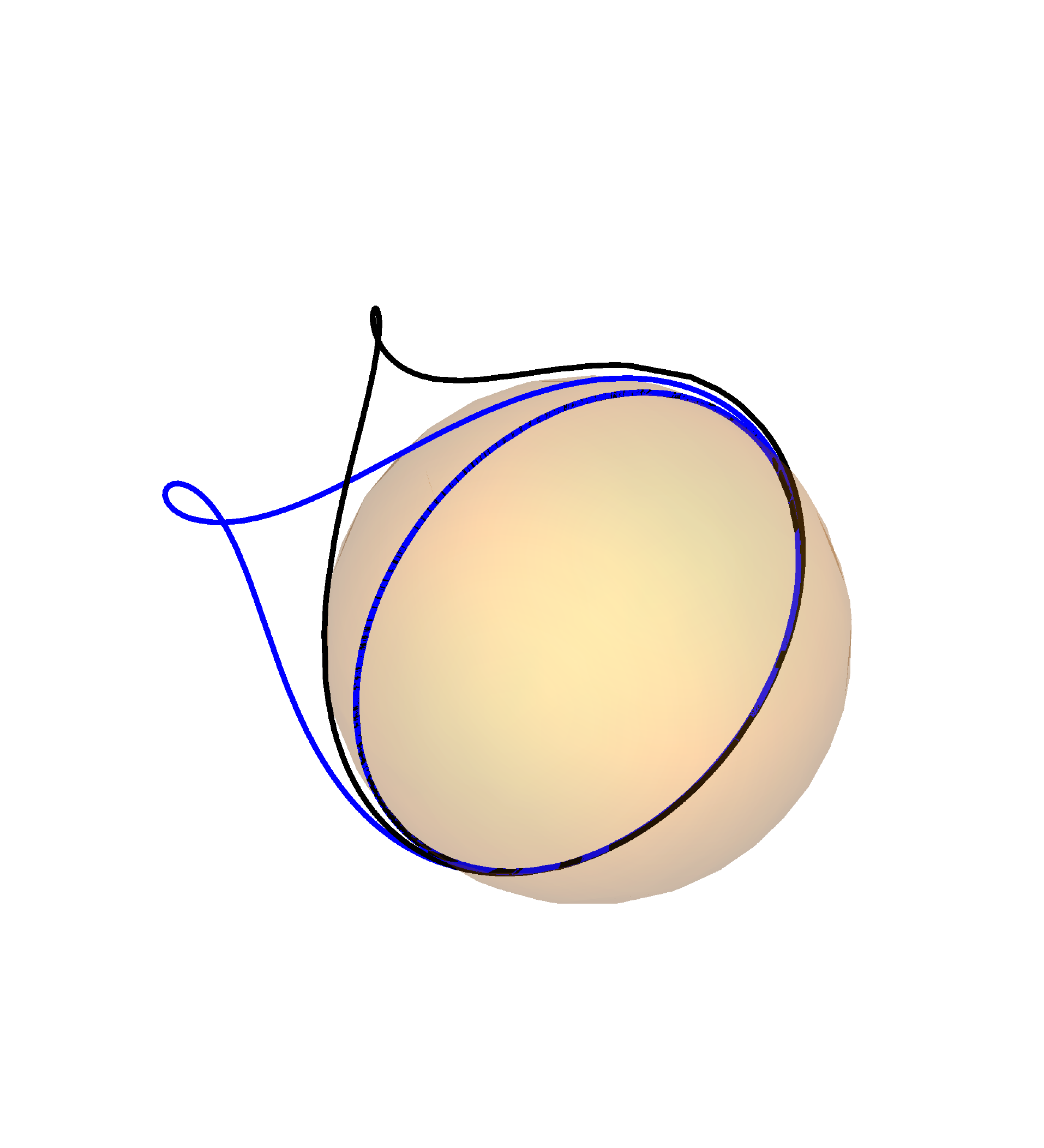}{fig3}
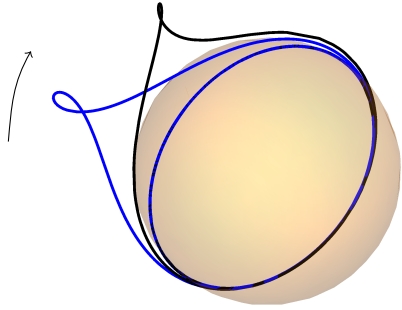
\begin{figure}
\begin{center}
\begin{tikzpicture}[scale=1]
\pgftext[at=\pgfpoint{0cm}{0cm},left,base]{\pgfuseimage{fig3}} 
\draw[->] (0.5,3.8) to[out=85,in=-120] (0.8,5);
\end{tikzpicture}
\caption{\footnotesize Stereographic image of a giant magnon solution of the lambda model for 2 nearby times (black and blue). Note that the solution appears as a kink on a string that is wound infinitely around a circle.}
\label{f4}
\end{center}
\end{figure} 

\subsection{Charges and mass shell relation}

The conserved charges carried by the solitons can be extracted from the monodromy \eqref{mon}. To this end, from the explicit form of the dressing transformation,
\EQ{
\Theta^{-1}(x=-\infty;z)\Theta(x=\infty;z)=\text{diag}\Big(\frac{z^2-\xi}{z^2-\xi^*},\frac{z^2+\xi^*}{z^2+\xi},\frac{z^2-\xi^*}{z^2-\xi},\frac{z^2+\xi}{z^2+\xi^*}\Big)^{s_\alpha}\ ,
\label{kcha}
}
where we have defined $s_\alpha=\text{sign}(\sin\alpha)$.
The $\pm1$ power in the above, accounts for the fact that when $\sin\alpha$ changes sign from $>0$ to $<0$ the asymptotic regimes $x=\pm\infty$ are swapped.
Hence the log of the monodromy is
\EQ{
\log {\cal W}(z)&=\lim_{x\to\infty}2 s_\alpha\, x(z^{-2}-z^{2})\Lambda\\ &+s_\alpha\text{diag}\Big(\log\frac{z^2-\xi}{z^2-\xi^*},\log\frac{z^2+\xi^*}{z^2+\xi},\log\frac{z^2-\xi^*}{z^2-\xi},\log\frac{z^2+\xi}{z^2+\xi^*}\Big)\ ,
}
Note that the divergent piece in the above does not contribute to the energy or momentum since $\STr(\Lambda\Lambda)=0$.  

The physical energy and momentum follow from the definitions \eqref{iuu}. The magnons also carry an additional abelian charge ${\cal Q}$. This is to be expected and corresponds to the dyonic  generalization of the giant magnon. The three conserved charges are
\EQ{
&{\cal E}^{\text{(sol)}}=\frac {k s_\alpha}{4\pi i}(\lambda^{-1}+\lambda)\log\left[\frac{\lambda^{-1}-\xi}{\lambda^{-1}-\xi^*}\cdot\frac{\lambda^{-1}+\xi^*}{\lambda^{-1}+\xi}
\cdot \frac{\lambda-\xi^*}{\lambda-\xi}\cdot\frac{\lambda+\xi}{\lambda+\xi^*}\right]\ ,\\[5pt]
&{\cal P}^{\text{(sol)}}=\frac {k s_\alpha}{4\pi i}(\lambda^{-1}-\lambda)\log\left[\frac{\lambda^{-1}-\xi}{\lambda^{-1}-\xi^*}\cdot\frac{\lambda^{-1}+\xi^*}{\lambda^{-1}+\xi}
\cdot \frac{\lambda-\xi}{\lambda-\xi^*}\cdot\frac{\lambda+\xi^*}{\lambda+\xi}\right]\ ,\\[5pt]
&{\cal Q}=\frac{k}{2\pi i}\log\left[\frac{\lambda^{-2}-\xi^2}{\lambda^{-2}-\xi^{*2}}\cdot\frac{\lambda^2-\xi^{*2}}{\lambda^2-\xi^2}\right]\ .
\label{kop}
}
In these expressions the branch of the logs must be chosen appropriately.

It is remarkable that these charges satisfy the dispersion relation \eqref{dis}. This proves that the dispersion relation holds at the classical and quantum level and therefore is not subject to quantum corrections (up to the possible shifts in the level $k$ mentioned earlier). In fact, one can explicitly relate the kinematic parameters $x^\pm$ used in the context of the S-matrix with the kinematic parameters $\xi$ and $\xi^\ast$ of the solitons simply as follows
\EQ{
&\text{magnon branch}:\qquad x^+= \frac{\lambda+\xi}{\lambda-\xi}\,,\\[5pt]
&\text{soliton branch}:\qquad\;\; x^+= \frac{\lambda-\xi}{\lambda+\xi}\,,
\label{themap}
}
with $x^-=(x^+)^\ast$ to ensure that the energy and momentum take real values.
Notice that the relationship between the definition in the magnon and the soliton branches is in agreement with~\eqref{mapping}. This leads to the following identification between the charges carried by the solitons and the (physical) energy and momentum of the S-matrix theory
\EQ{
{\cal Q}= a\,,\qquad
{\cal E}^{\text{(phys)}} = s_\alpha\,  {\cal E}^{\text{(sol)}}\,,\qquad
{\cal P}^{\text{(phys)}}= -s_\alpha\, {\cal P}^{\text{(sol)}}\,,
\label{Corr}
}
which shows that the bosonic $\SU(4)/\Sp(4)$ solitons reproduce the quantum spectrum of the $q$-deformed S-matrix.

The first equation in~\eqref{Corr} associates the soliton to the representation $\langle a-1,0\rangle$ by means of the identification of 
the soliton charge ${\cal Q}$ with the integer~$a$. This equation is a quantization condition that could also be deduced in the semiclassical limit by means of the Bohr-Sommerfeld approach.

Once the charge ${\cal Q}$ is quantized, the expression for ${\cal Q}$ given by \eqref{kop} implicitly determines $\alpha$ as a function of $\theta$. Although there are several branches of solutions, only the one with $s_\alpha\,  {\cal E}^{\text{(sol)}}>0$ matches the quantum spectrum. This is a similar phenomenon to a free field where classically there are modes with ${\cal E}=\pm\sqrt{{\cal P}^2+M^2}$ but it is only the positive energy modes that match the spectrum of quantum states.
In our case, $s_\alpha<0$ and $s_\alpha>0$ in the magnon and soliton branches, respectively and, by means of~\eqref{themap}, both correspond to $\text{Im\,}(x^+)>0$. Notice that the soliton energy provided by the monodromy ${\cal E}^{\text{(sol)}}$ turns out to be negative in the magnon branch. However, we will show in the next paragraph that  ${\cal Q}$ and ${\cal E}^{\text{(phys)}}=s_\alpha\,  {\cal E}^{\text{(sol)}}$ (not ${\cal E}^{\text{(sol)}}$ by itself) determine the asymptotic values of the lambda model field $\CF$.

In the sigma model limit, the mass shell condition reduces to that of the dyonic giant magnon \eqref{yr2} \cite{Chen:2006gea,Dorey:2006dq}. The solution has non-vanishing momentum and so in the sigma model limit a giant magnon must be put together with other giant magnons to ensure the periodicity condition on the total momentum ${\cal P}=0$.
 If we further take the limit of large 't~Hooft coupling, then \eqref{yr2} reduces to the relativistic mass shell condition ${\cal E}^2={\cal Q}^2+{\cal P}^2$ which is valid in the string theory on the plane wave limit of AdS$_ 5{\times} S^5$ \cite{Berenstein:2003gb}.
 
The true nature of the lambda model solitons (giant magnons) emerges when looking at the form of the group valued field $\CF$: they are kinks that interpolate between different vacuum solutions.
To spell this out, recall that, as explained in section~\ref{s1}, each solution of the associated linear system gives rise to lambda model field configurations 
of the form
\EQ{
\CF(x)= \Psi(x;\lambda^{1/2})\, V(\lambda)\, \Psi^{-1}(x;\lambda^{-1/2})\,,
}
where $V(\lambda)\in F$ is constant and arbitrary.
Let us consider the gauged fixed field configurations corresponding to 
a $\lambda$ independent group element of the form
\EQ{
V(\lambda)= \exp\left(\alpha\,\Sigma_3^{(3)} + \beta\,\Sigma_3^{(4)} \right) \in H=\SU(2)_3\times \SU(2)_4\,,
}
where
\EQ{
\Sigma_3^{(3)}=i\,\text{diag}(1,-1,0,0)\in \msu(2)_3\,,\qquad
\Sigma_3^{(4)}=i\, \text{diag}(0,0,1,-1)\in \msu(2)_4\,.
}
Then, for $s_\alpha>0$,
\EQ{
&
{\cal F}(t,+\infty)={\cal F}_0(t,+\infty)\,\exp\left(-\frac{2\pi}{k}\, \frac{{\cal E}^{\text{(phys)}}}{\lambda^{-1}+\lambda}\,\Lambda_2\right)\,
\exp\bigg(\alpha\,\Sigma_3^{(3)}+
\left(\beta +\frac{\pi}{k}\, {\cal Q}\right)\,\Sigma_3^{(4)}\bigg)\,,\\[5pt]
&
{\cal F}(t,-\infty)={\cal F}_0(t,-\infty)\,\exp\left(\frac{2\pi}{k}\, \frac{{\cal E}^{\text{(phys)}}}{\lambda^{-1}+\lambda}\,\Lambda_2\right)\,
\exp\bigg(\left(\alpha+\frac{\pi}{k}\, {\cal Q}\right)\,\Sigma_3^{(3)}+ \beta\,\Sigma_3^{(4)}\bigg)\,,
\label{VacInt}
}
where we have taken into account the compensating scalar factor~\eqref{Sfactor}.
These asymptotic values are swapped when $s_\alpha$ changes from $>0$ to $<0$.

The lambda model action~\eqref{dWZW} includes a Wess-Zumino topological term whose consistency imposes two types of quantization conditions. The first one is the well known quantization of the coupling constant, whose role is taken by the level~$k$ of the super WZW part of the action. The second is a quantization condition on the boundary conditions that can be considered in the decompactification limit, which is required to define the WZ term on a world-sheet with boundary (see~\cite{NA-kinks} and references therein). 
In our case, this condition applies to the boundary conditions taking values in $H$.

Our gauge fixing conditions leave a residual symmetry under global (vector) gauge transformations
\EQ{
\CF(x) \to U \CF(x) U^\dagger \,,\qquad U\in H\,.
} 
This shows that, on the boundary $x=\pm\infty$, the field $\CF$ actually takes values in conjugacy classes, or co-adjoint orbits, of $H$.
Then, following~~\cite{NA-kinks}, the consistency of the WZ term requires that
\EQ{
|\alpha| = \frac{2\pi}{k}\, j_1\,,\qquad |\beta| = \frac{2\pi}{k}\, j_2\,,\qquad j_1,\,j_2 <\frac{k}2\,,
}
where $j_1$ and $j_2$ are $\msu(2)$ spins. Since one can change $\Sigma_3^{(3/4)}\to -\Sigma_3^{(3/4)}$ by means of a conjugation under $H$, this gives rise to four non-equivalent allowed boundary conditions that can be labeled as
\EQ{
&
\Big[\underbrace{j_1+\frac{1}2{\cal Q},\, j_2}_{x\to-\infty}\, \Big| 
\underbrace{j_1,\, j_2+\frac{1}2{\cal Q}}_{x\to+\infty}\Big]\,,\qquad
\Big[j_1-\frac{1}2{\cal Q},\, j_2\, \Big| j_1,\, j_2+\frac{1}2{\cal Q}\Big]\,,\\[7pt]
&\Big[j_1+\frac{1}2{\cal Q},\, j_2\, \Big| j_1,\, j_2-\frac{1}2{\cal Q}\Big]\,,\qquad
\Big[j_1-\frac{1}2{\cal Q},\, j_2\, \Big| j_1,\, j_2-\frac{1}2{\cal Q}\Big]\,.
\label{bcc}
}
This confirms the kink nature of the giant magnons (solitons) of the lambda model. In addition, it provides an alternative interpretation of the quantization rule ${\cal Q}=a$ in~\eqref{Corr}, where~$a$ is a positive integer number $<k$.

\subsection{Classical giant magnon scattering}\label{s3.3}

In this section, we consider the scattering of giant magnons from a classical perspective. 
The scattering of classical solitons in integrable theories can be described as a time delay experienced by one of the solitons as the other passes through it as we illustrate in fig.~\ref{f5}.
The time delay experienced by giant magnon 1 as giant magnon 2 moves through from $x=+\infty$ to $x=-\infty$ can equally well be described as a shift in giant magnons 1's position of 
\EQ{
\Delta x_0=-\sinh\theta_1\Delta t
\label{kkz}
} 
in its rest frame. We now turn to a calculation of the shift $\Delta x_0$.

\FIG{
\begin{tikzpicture} [scale=1]
\filldraw[blue!20] (2.1,2.05) circle (1cm);
\draw[-] (0,0) -- +(40:4.4cm);
\draw[-] (0.8,1.3) -- +(40:4cm);
\draw[-] (3.5,0) -- +(130:4cm);
\draw[-] (3.2,1.3) -- +(130:4cm);
\draw[->,thick] (0,0.05) .. controls (2.5,2.1) and (1.4,1.8) .. (3.9,3.85);
\draw[->,thick] (3.5,0.05) .. controls (1.2,2.8) and (2.8,1.8) .. (0.65,4.3);
\draw (-0.6,0) .. controls (-0.2,0) and (-0.1,1.2) .. (0,1.2) .. controls (0.1,1.2) and (0.4,0) .. (0.8,0);
\draw[densely dashed] (0,-0.1) -- (0,1.4);
\begin{scope}[xshift=3.5cm,yshift=0cm]
\draw (-0.6,0) .. controls (-0.2,0) and (-0.1,0.9) .. (0,0.9) .. controls (0.1,0.9) and (0.4,0) .. (0.8,0);
\draw[densely dashed] (0,-0.1) -- (0,1.1);
\end{scope}
\begin{scope}[xshift=3.9cm,yshift=3.8cm]
\draw (-0.6,0) .. controls (-0.2,0) and (-0.1,1.2) .. (0,1.2) .. controls (0.1,1.2) and (0.4,0) .. (0.8,0);
\draw[densely dashed] (0,-0.1) -- (0,1.4);
\end{scope}
\begin{scope}[xshift=0.65cm,yshift=4.3cm]
\draw (-0.6,0) .. controls (-0.2,0) and (-0.1,0.9) .. (0,0.9) .. controls (0.1,0.9) and (0.4,0) .. (0.8,0);
\draw[densely dashed] (0,-0.1) -- (0,1.1);
\end{scope}
\draw[->]  (-3,2) -- (-2,2);
\draw[->]  (-3,2) -- (-3,3);
\node at (-1.7,2) (a2) {$x$};
\node at (-3,3.3) (a3) {$t$};
\draw[densely dashed] (3,2) -- (3,3.5);
\draw[densely dashed] (3,2.5) -- (6,2.5);
\draw[densely dashed] (3,3.1) -- (6,3.1);
\draw[->] (5.5,2.1) -- (5.5,2.5);
\draw[->] (5.5,3.5) -- (5.5,3.1);
\node at (6.3,2.8) (a1) {$\Delta t$};
\end{tikzpicture}\caption{\footnotesize The scattering of two giant magnons in space-time. The giant magnons scatter and retain their shape and velocities. The only effect of the scattering is to introduce a time delay on the motion of the giant magnons as shown. Note that an attractive force between the giant magnons produces a time advance $\Delta t<0$, whereas, here we have illustrated a repulsive force $\Delta t>0$.}
\label{f5}
}

If one reviews the construction of the one giant magnon solution, briefly summarized earlier in this section, one sees that the 
position of the giant magnon is neatly encoded in the scalar quantity
\EQ{
\BF_1^*\cdot\BF_1=e^{X+X^*}+e^{-X-X^*}=e^{2x\sin\alpha}+e^{-2x\sin\alpha}\ ,
}
in the rest frame.

Now consider the situation in which there are two giant magnons. Our strategy is to work in the rest frame 
of giant magnon 1 and then calculate the effect on it as giant magnon 2 travels through it from positive to negative $x$. The final result can then be boosted to an arbitrary frame.
The dressing method gives an elegant way of studying the resulting scattering event \cite{Hollowood:2009tw}; in this paradigm we think of giant magnon $1$ as dressing giant magnon $2$; in other words, we construct the two giant magnon solution by a two stage process:
\EQ{
\Psi_0\overset{\Theta^{(2)}}\longrightarrow\Psi_2\overset{\Theta^{(1)}}\longrightarrow\Psi_1\ .
}
As discussed above, the spacetime position of giant magnon $1$ is 
encoded in the vector $\BF^{(1)}_1$ which is now given by the dressed quantity
\EQ{
\BF^{(1)}_1=\Psi_2(\sqrt{\xi_1^*})\Bvarpi^{(1)}=\Theta^{(2)}(\sqrt{\xi_1^*})\, \Psi_0(\sqrt{\xi_1^*})\,\Bvarpi^{(1)}\ .
}
Without-loss-of-generality, we can fix the internal orientation of the collective coordinates of giant magnon 1 to be
\EQ{
\Bvarpi^{(1)}=(1,0,1,0)\ .
}
Magnon 2 then has a general orientation in $S^2\times S^2$ encoded in the vector 
\EQ{
\Bvarpi^{(2)}=(c_1,c_2,c'_1,c'_2)\ ,
\label{xff}
}
with redundancy 
$(c_1,c_2)\sim \zeta(c_1,c_2)$ for $\zeta\in\mathbb C$ and similarly for $(c_1',c_2')$.

In order to extract the shift in position of giant magnon 1 as giant magnon 2 passes we need the
 asymptotic limits 
\EQ{
\Theta^{(2)}_\pm(z)=\lim_{x\to\pm\infty}\Theta^{(2)}(z)\ .
}
The latter are given by
\EQ{
\Theta^{(2)}_\pm(z)=1+\frac{\sigma_i(\Bvarpi^{(2)}_\pm)\Gamma^{(\pm)-1}_{ij}\sigma_j(\Bvarpi^{(2)}_\pm)^\dagger}{z^2-\sigma_j(\xi_2)}\ ,
}
where
\EQ{
\Bvarpi^{(2)}_+=(c_1,c_2,0,0)\ ,\qquad \Bvarpi^{(2)}_-=(0,0,c_1',c_2')\ ,
}
and where we have defined the $4\times 4$ matrices
\EQ{
\Gamma^{(\pm)}_{ij}=\frac{\sigma_i(\Bvarpi^{(2)}_\pm)^*\cdot\sigma_j(\Bvarpi^{(2)}_\pm)}{\sigma_i(\xi_2)-\sigma_j(\xi_2)^*}\ .
}

The strategy is to now calculate $\BF^{(1)*}_1\cdot\BF^{(1)}_1$ for the giant magnon 1 in the two asymptotic regimes for giant magnon 2.
For the $+$ region, when giant magnon 2 is well to the right of giant magnon 1 we find
\EQ{
\BF^{(1)*}_1\cdot\BF^{(1)}_1=e^{2x \sin\alpha_1}+\Big(|c_1|^2\left|\frac{\xi_1-\xi_2}{\xi_1-\xi_2^{*}}\right|^2
+|c_2|^2\left|\frac{\xi_1+\xi_2^{*}}{\xi_1+\xi_2}\right|^2\Big)e^{-2x \sin\alpha_1}\ ,
}
while in the $-$ region, when giant magnon 2 is well to the left of giant magnon 1 one takes the above expression and replaces $x\to-x$ along with $c_i\to c'_i$. 

The shift in the position of the giant magnon 1 in its rest frame caused by the interaction is then
\EQ{
\Delta x_0&=-\frac{1}{4\sin\alpha_1}\log\Big[\Big(|c_1|^2\left|\frac{\xi_1-\xi_2}{\xi_1-\xi_2^{*}}\right|^2
+|c_2|^2\left|\frac{\xi_1+\xi_2^{*}}{\xi_1+\xi_2}\right|^2\Big)\\ &~~~~~~~~~~~~~\qquad\qquad\times\Big(|c'_1|^2\left|\frac{\xi_1-\xi_2}{\xi_1-\xi_2^{*}}\right|^2
+|c'_2|^2\left|\frac{\xi_1+\xi_2^{*}}{\xi_1+\xi_2}\right|^2\Big)\Big]\ .
}
The corresponding time delay in an arbitrary frame is then
\EQ{
\Delta t=-\frac{\Delta x_0}{\sinh\theta_1}\ .
}
Note that for the cases of interest the time delay $\Delta t$ is actually negative, i.e.~it is a time advance. This formulae will  be the basis of a semi-classical test of the quantum S-matrix in the following sections.

\section{S-matrix and semi-classical limit}\label{s5}

In this section, we propose that the quantum scattering of giant magnons in the lambda theory is described by the S-matrix constructed in a series of papers \cite{Hoare:2011nd,Hoare:2011wr,Hoare:2012fc,Hoare:2013hhh} based on the solution of the Yang-Baxter equation constructed by Beisert and Koroteev \cite{Beisert:2008tw}. This S-matrix can be viewed as a deformation of the AdS$_ 5{\times} S^5$ giant magnon S-matrix\footnote{This S-matrix was determined in \cite{Beisert:2005tm} and the all-important dressing phase in \cite{Dorey:2007xn,Gromov:2007cd,Arutyunov:2009kf,Volin:2009uv,Kruczenski:2009kc,Vieira:2010kb}.} where the Yangian invariance is deformed into a quantum group with a quantum parameter $q=\exp(i\pi/k)$. The S-matrix respects the symmetry that remains around the vacuum solution. This subgroup includes $\SU(2|2){\times}\SU(2|2)$, which becomes enhanced to the Yangian in the sigma model and the quantum group in the lambda model.

The question is how in detail are the quantum states of the giant magnon related to the classical solution? The first point to make is that the classical solution has an internal collective coordinate a complex 8-vector, on which the symmetry group $\SU(2|2){\times}\SU(2|2)$ acts. Let us concentrate on the solutions without Grassmann modes turned on. In that case, the classical solution has a complex 4-vector $\Bvarpi$: referring to fig.~\ref{f1}, $\SU(2)_3$ acts on the first two elements and $\SU(2)_4$ on the last two:
\EQ{
\Bvarpi=(c_1,c_2,c_1',c_2')\ .
}
Up to shifts in the space time coordinates, the solution is invariant under the re-scalings $(c_1,c_2)\to\zeta(c_1,c_2)$ and $(c_1',c_2')\to\zeta'(c_1',c_2')$ and so the bosonic giant magnon carries an internal moduli space $S^2\times S^2$ on which $\SU(2)_3\times\SU(2)_4$ has a natural action.

The quantum states of the giant magnons should correspond to the states of spin $a/2$ in each of $\SU(2)_3$ and $\SU(2)_4$. The relation between quantum states and classical configurations in the correspondence limit, i.e.~large $a$, is a familiar one. The Hilbert space contains many more states than the classical system. However, classical states should correspond to quasi-classical, or coherent, states. These states are obtained by acting on the highest weight state by the action of $\SU(2)$ on the Hilbert space $U\ket{j=a/2,m=a/2}$, $U\in\SU(2)$: so the states with maximal spin along any direction on $S^2$. These states are labelled by a point on $S^2\times S^2$ matching precisely the classical configurations.

If we think of the giant magnon state with abelian charge $a$ as being a bound state of $a$ fundamental giant magnons transforming in the $j=\frac12$ representation, then the coherent states in the bound state correspond to 
\EQ{
\ket{\Psi}=\big(c_1\ket{\uparrow}+c_2\ket{\downarrow}\big)^{\otimes a}\otimes
\big(c'_1\ket{\uparrow}+c'_2\ket{\downarrow}\big)^{\otimes a}\ .
\label{sdp}
}

In the lambda model, the group symmetry is deformed into a quantum group: so each $\msu(2)\to U_q(\msu(2))$ with quantum parameter $q=\exp(i\pi/k)$. The representation structure is largely similar to the undeformed group \cite{Hoare:2011nd}. However, since $q$ is a root of unity, there is a truncation of the states to $j\leq\frac k2-1$. Importantly, however, 
the states are realized in the IRF or RSOS version of the quantum group \cite{Hoare:2013hhh}. In this picture, states are kinks that interpolate between a set of vacua and the kink Hilbert space is much more restricted compared with the original Hilbert space. For each $U_q(\msu(2))$, the vacua are associated to 
the set of representations with spins in the set $\{0,\frac12,1,\frac32,\ldots,\frac k2-1\}$. The basic $a=1$ giant magnons correspond to kinks $K^{j_1,j_2}_{j_3,j_4}$ with $j_1=j_2\pm\frac12$ and $j_3=j_4\pm\frac12$ which are identified in the original ``vertex picture" with the states $\ket{\uparrow\uparrow}$,  $\ket{\uparrow\downarrow}$,  $\ket{\downarrow\uparrow}$ and  $\ket{\downarrow\downarrow}$. 

The analogue of the coherent states \eqref{sdp} in the kink Hilbert space are the states
\EQ{
K^{j_1+\frac12a,j_1}_{j_2+\frac12a,j_2}\ ,\qquad
 K^{j_1+\frac12a,j_1}_{j_2-\frac12a,j_2}\ ,\qquad K^{j_1-\frac12a,j_1}_{j_2+\frac12a,j_2}\ ,\qquad
 K^{j_1-\frac12a,j_1}_{j_2-\frac12a,j_2}\ .
\label{sdp2}
}
These quantum kinks are clearly related to the boundary conditions in \eqref{bcc}.
In terms of the classical soliton solutions, the quantum states  correspond to solitons with internal collective coordinates
\EQ{
&\Bvarpi_{\uparrow\uparrow}=(1,0,1,0)\ ,\qquad\Bvarpi_{\uparrow\downarrow}=(1,0,0,1)\ ,\\
&\Bvarpi_{\downarrow\uparrow}=(0,1,1,0)\ ,\qquad\Bvarpi_{\downarrow\downarrow}=(0,1,0,1)\ .
\label{num1}
}
It is the scattering of these states that we will match to the classical scattering theory.

\subsection{The S-matrix and bound state scattering}\label{s5.3}

Although the spectrum of giant magnon bound states is captured exactly at the semi-classical level, we only expect the S-matrix of such states to match with the classical scattering theory of giant magnons in the semi-classical limit $k\to\infty$ and $g\to\infty$ with fixed $g/k$, i.e.~fixed $\lambda$. For the scattering theory, the semi-classical states are those with abelian charge $a\to\infty$ with $a/k$ fixed.

Now we turn to the S-matrix. It is defined in terms of the scattering of the 16 basic states with $a=1$.
Scattering of bound states is then determined by applying the bootstrap principle. Before we describe this procedure, we need, first of all, to describe the various kinematical variables that can be used to label states.

We start with the parameters that appear in the classical dressing method, these are 
\EQ{
\xi=e^{-\theta-i\alpha}\ ,\qquad \xi^*=e^{-\theta+i\alpha}\ ,
}
where $\theta$ is the rapidity and $\alpha=\alpha(\theta,a)$ obtained by fixing ${\cal Q}=a$ and the rapidity in \eqref{kop} and solving for $\alpha$. 

As we described in section \ref{s2}, the S-matrix is usually presented in terms of pair of 
variables $x^\pm$ which are related to the soliton parameters $\xi$ and $\xi^*$ by the map \eqref{themap}.\footnote{In the following we will work on the magnon branch for simplicity although there is no fundamental obstruction in applying the same formalism to the soliton branch.}
Another convenient kinematic variable is the {\it pseudo rapidity\/} defined
by 
\EQ{
e^{4\nu}=\frac{1-\lambda^2\xi^2}{\lambda^2-\xi^2}\cdot\frac{1-\lambda^2\xi^{*2}}{\lambda^2-\xi^{*2}}\ .
}
Note that in the relativistic limit, $\lambda\to0$, the pseudo rapidity becomes equal to the ordinary rapidity $\nu=\theta$.

The relation between the pseudo rapidity and the $x^\pm$ variables is best understood in terms 
of a map $x(\nu)$,
\EQ{
x+\frac1x=\frac{2}{\lambda^{-2}-\lambda^2}\big(\lambda e^{2\nu}-\lambda^{-1}\big)\ .
\label{j22}
}
So $\nu$ is naturally valued on a cylinder $\nu\sim\nu+i\pi$. The map $x(\nu)$ is branched at $\pm\log\lambda$ and we define $\EuScript C$ to be the branch cut.
The pseudo rapidity determines the pair $x^\pm$ via
\EQ{
x^\pm=x\Big(\nu\pm\frac{i\pi a}{2k}\Big)\ .
\label{j23}
}

The S-matrix is usually expressed as a function $S(x^\pm_1,x^\pm_2)$ and the $a$ bound state is formed by putting together $a$ basic states with parameters as
\EQ{
x_1^+=x_2^-\ ,\quad x_2^+=x_3^-\ ,\ \ldots\ ,\quad x_{a-1}^+=x_a^-\ ,
\label{p12}
}
so that the kinematic variables of the bound state are $x_B^+=x_a^+$ and $x_B^-=x_1^-$. 
The structure of bound states is particularly simple in terms of the 
pseudo rapidity. If $\nu$ is the pseudo rapidity of the bound state then its constituents have
\EQ{
\nu_j=\nu-\frac{i\pi}{2k}(a+1-2j)\ ,
\label{etas}
}
$j=1,2,\ldots,a$.\footnote{On the soliton branch in the  relativistic limit, these pseudo rapidities correspond to  relativistic rapidities
$\theta_j=\theta+\frac{i\pi}{2k}(a+1-2j)$.}

The bootstrap equations determine the S-matrix elements of the bound states in terms of the those of the basic states. The equations are represented pictorially in fig.~\ref{f6}
\FIG{
\begin{tikzpicture} [scale=0.8]
\draw[->,thick,double] (-1,0) -- (1,2);
\draw[->,thick,double] (1,0) -- (-1,2);
\draw[thick] (-1,0) -- +(-165:2cm);
\draw[thick] (-1,0) -- +(-155:2cm);
\draw[thick] (-1,0) -- +(-145:2cm);
\draw[thick] (-1,0) -- +(-135:2cm);
\draw[thick] (-1,0) -- +(-125:2cm);
\draw[thick] (-1,0) -- +(-115:2cm);
\draw[thick] (-1,0) -- +(-105:2cm);
\draw[thick] (1,0) -- +(-66:2cm);
\draw[thick] (1,0) -- +(-52:2cm);
\draw[thick] (1,0) -- +(-38:2cm);
\draw[thick] (1,0) -- +(-24:2cm);
\filldraw[black] (-1,0) circle (4pt);
\filldraw[black] (1,0) circle (4pt);
\filldraw[black](0,1) circle (6pt);
\node at (4.3,0.2) (a1) {$\displaystyle =\sum_\text{internal lines}$};
\begin{scope}[xshift=8cm,yshift=-0.5cm]
\draw[thick] (1,2) -- +(-165:3cm);
\draw[thick] (1,2) -- +(-155:3.1cm);
\draw[thick] (1,2) -- +(-145:3.2cm);
\draw[thick] (1,2) -- +(-135:3.4cm);
\draw[thick] (1,2) -- +(-125:3.6cm);
\draw[thick] (1,2) -- +(-115:3.8cm);
\draw[thick] (1,2) -- +(-105:4cm);
\draw[thick] (-1,2) -- +(-66:4cm);
\draw[thick] (-1,2) -- +(-52:3.8cm);
\draw[thick] (-1,2) -- +(-38:3.6cm);
\draw[thick] (-1,2) -- +(-24:3.4cm);
\draw[->,thick,double] (-1,2) -- +(135:1.2);
\draw[->,thick,double] (1,2) -- +(45:1.2);
\filldraw[black] (-1,2) circle (4pt);
\filldraw[black] (1,2) circle (4pt);
\end{scope}
\end{tikzpicture}
\caption{\footnotesize A pictoral representation of the the bootstrap/fusion equations for the case $a_1=7$ and $a_2=4$. In general one has the sum over all the possible quantum numbers of the internal lines on the right-hand side. Our focus is on scalar processes for which the quantum numbers are fixed and no sum is necessary.}
\label{f6}
}
What makes the bootstrap equations difficult to apply is that one has to sum over the quantum numbers of the states on the internal lines. However, if we choose the external states appropriately, the internal states are fixed uniquely and the bootstrap equations are trivialized. In this case, 
the bootstrap equation that gives the scattering of bound states $a_1$ and $a_2$ with pseudo rapidities $\nu_1$ and $\nu_2$ is just a simple product
\EQ{
S_{a_1a_2}=\prod_{j=1}^{a_1}\prod_{l=1}^{a_2}S\Big(\nu_1+\frac{i\pi}{2k}(a_1-2j+1),\nu_2+\frac{i\pi}{2k}(a_2-2l+1)\Big)\ .
\label{boeq}
}
where $S(\nu_1,\nu_2)$ is the S-matrix element of the constituent states written in terms of the pseudo rapidity.

The external states that have scalar bootstrap equations are precisely the states that only involve the up and down states $\ket{\uparrow\uparrow}$, $\ket{\uparrow\downarrow}$, $\ket{\downarrow\uparrow}$ and $\ket{\downarrow\downarrow}$. This is particularly convenient because these are also the states that lie in the kink Hilbert space of the lambda model.

The scattering of these states involves essentially three inequivalent scattering processes:
\EQ{
S^{(1)}=\begin{tikzpicture}[baseline=-0.65ex,scale=0.9]
    \draw[->,thick,double] (-1,-1) -- (1,1);
    \draw[<-,thick,double] (-1,1) -- (1,-1);
\filldraw[black] (0,0) circle (6pt);        
  \node at (-1.3,1.3) (a1) {$\uparrow\uparrow$};
       \node at (1.3,-1.3) (a2) {$\uparrow\uparrow$};
           \node at (1.3,1.3) (a3) {$\uparrow\uparrow$};
    \node at (-1.3,-1.3) (a4) {$\uparrow\uparrow$};
   \node at (-1,0) (a1) {$j+1\atop j'+1$};
       \node at (0,-1) (a2) {$j+\frac12\atop j'+\frac12$};
           \node at (1,0) (a3) {$j\atop j'$};
    \node at (0,1) (a4) {$j+\frac12\atop j+\frac12$};   
\end{tikzpicture}\ \qquad
S^{(2)}=\begin{tikzpicture}[baseline=-0.65ex,scale=0.9]
    \draw[->,thick,double] (-1,-1) -- (1,1);
    \draw[<-,thick,double] (-1,1) -- (1,-1);
\filldraw[black] (0,0) circle (6pt);  
  \node at (-1.3,1.3) (a1) {$\uparrow\downarrow$};
       \node at (1.3,-1.3) (a2) {$\uparrow\downarrow$};
           \node at (1.3,1.3) (a3) {$\uparrow\uparrow$};
    \node at (-1.3,-1.3) (a4) {$\uparrow\uparrow$};
   \node at (-1,0) (a1) {$j+1\atop j'$};
       \node at (0,-1) (a2) {$j+\frac12\atop j'-\frac12$};
           \node at (1,0) (a3) {$j\atop j'$};
    \node at (0,1) (a4) {$j+\frac12\atop j+\frac12$};   
\end{tikzpicture}\ \qquad
S^{(3)}=\begin{tikzpicture}[baseline=-0.65ex,scale=0.9]
     \draw[->,thick,double] (-1,-1) -- (1,1);
    \draw[<-,thick,double] (-1,1) -- (1,-1);
\filldraw[black] (0,0) circle (6pt);  
  \node at (-1.3,1.3) (a1) {$\downarrow\downarrow$};
       \node at (1.3,-1.3) (a2) {$\downarrow\downarrow$};
           \node at (1.3,1.3) (a3) {$\uparrow\uparrow$};
    \node at (-1.3,-1.3) (a4) {$\uparrow\uparrow$};
   \node at (-1,0) (a1) {$j\atop j'$};
       \node at (0,-1) (a2) {$j-\frac12\atop j'-\frac12$};
           \node at (1,0) (a3) {$j\atop j'$};
    \node at (0,1) (a4) {$j+\frac12\atop j+\frac12$};   
\end{tikzpicture}
\label{wea2}
}
where
\EQ{
S^{(1)}=\frac1{\sigma_{12}^2}\cdot\frac{x_1^+x_2^-}{x_1^-x_2^+}\cdot\frac{x_1^--x_2^+}{x_1^+-x_2^-}\cdot
\frac{1-\frac1{x_1^-x_2^+}}{1-\frac1{x_1^+x_2^-}}\ .
\label{xpp}
}
In the above, $\sigma_{12}$ is the $q$-deformed version of the dressing phase \cite{Hoare:2011wr} which reproduces the dressing factor of the string S-matrix in the appropriate limit \cite{Volin:2009uv,Vieira:2010kb,Arutyunov:2009kf}. The other elements are given by $S^{(2)}=f_{12}S^{(1)}$, $S^{(3)}=f_{12}S^{(2)}$ where
\EQ{
f_{12}=q^{-1}\frac{\sqrt{[2j+2][2j]}}{[2j+1]}\cdot\frac{x_1^+-x_2^+}{x_1^--x_2^+}\cdot
\frac{1-\frac1{x_1^+x_2^-}}{1-\frac1{x_1^-x_2^-}}
}
and
\EQ{
[n]=\frac{q^n-q^{-n}}{q-q^{-1}}\ .
\label{qnb}
}

These S-matrix elements were constructed in \cite{Hoare:2011wr,Hoare:2013hhh} and reproduce the elements of the S-matrix for the AdS$_ 5{\times} S^5$ case in the limit $k\to\infty$. 

In the relativistic limit, $\lambda\to0$, these S-matrix elements reduce to the familiar looking trigonometric expressions in the rapidity difference $\theta=\theta_1-\theta_2$:
\EQ{
S^{(1)}=\frac1{\sigma_{12}^2}\cdot\frac{\sinh(\frac\theta2+\frac{i\pi}{2k})}{\sinh(\frac\theta2-\frac{i\pi}{2k})}\cdot
\frac{\cosh(\frac\theta2+\frac{i\pi}{2k})}{\cosh(\frac\theta2-\frac{i\pi}{2k})}\ ,
}
along with
\EQ{
f_{12}=\frac{\sinh(\frac\theta2)}{\cosh(\frac\theta2)}\cdot
\frac{\cosh(\frac\theta2-\frac{i\pi}{2k})}{\sinh(\frac\theta2+\frac{i\pi}{2k})}\ .
}

\subsection{The Semi-Classical Limit}\label{s5.4}

The semi-classical limit involves taking $k\to\infty$ and $g\to\infty$ with $\lambda$ in \eqref{crel} fixed. In this limit, for the scattering of the $a=1$ basic states
\EQ{
x^\pm=x\Big(\nu\pm\frac{i\pi}{2k}\Big)\longrightarrow x(\nu)\pm\frac{i\pi}{2k}x'(\nu)+{\cal O}(k^{-2})\ ,
}
and we can expand the S-matrix as
\EQ{
S=\exp\Big[\frac{i\pi}kF+{\cal O}(k^{-2})\Big]\ .
}
Now we are in a position to apply the bootstrap equations \eqref{boeq} to find the scattering of the quasi-classical soliton states with $\hat a _1=\pi a_1/2k$ and $\hat a_2=\pi a_2/2k$ fixed as $k\to\infty$:
\EQ{
\JLMc{S_{a _1a _2}=} \exp\Big[\frac {i\pi}k\sum_{j=1}^{a_1}\sum_{l=1}^{a_2}F\Big(\nu_1+\frac{i\pi  }{2k}(a_1-2j+1),\nu_2+\frac{i\pi  }{2k}(a_2-2l+1)\Big)\Big]\ .
}
To leading order in $1/k$ we can replace the sums by integrals:
\EQ{
\sum_{j=1}^{a}g\Big(\frac{\pi  }{2k}(a-2j+1)\Big)\longrightarrow\frac{k }\pi\int_{-\hat a}^{\hat a} dv\, g(v)\ ,
}
to arrive at
\EQ{
S_{a_1a_2}=\exp\left[\frac{ik   }{\pi}\int_{-\hat a_1}^{\hat a_1}dv_1\int_{-\hat a_2}^{\hat a_2}dv_2\,F(\nu_1+iv_1,\nu_2+iv_2)\right]\ .
}
Writing $F(\nu_1,\nu_2)=-d\log G(\nu_1,\nu_2)/d\nu_2$ gives
\EQ{
S_{a_1a_2}=\exp\left[-\frac{k  }{\pi}\int_{-\hat a_1}^{\hat a_1}dv_1\,\log\frac{G(\nu_1+iv_1,\nu_2+i\hat a_2)}{G(\nu_1+iv_1,\nu_2-i\hat a_2)}\right]\ .
}

Before proceeding we have to specify which terms in the exponent above we need to keep track of when comparing with the classical time delays. The Jackiw-Woo formula \cite{Jackiw:1975im} that we use in due course is derived in quantum mechanics for a particle scattering off a potential and as such has been found to capture the semi-classical limit of the S-matrix of a relativistic QFT in $1+1$-dimensions. We will find that it continues to capture the terms
which are non-trivial functions of both $\theta_1$ and $\theta_2$ (or $\nu_1$ or $\nu_2$) 
in our non-relativistic field theory setting. However,  the S-matrix can also depend on multiplicative factors that are just functions of either $\theta_1$ or $\theta_2$ separately. Such terms can be interpreted as rapidity re-definitions of the one particle states and, consequently, we will not keep track of such terms.

Rather than computing the integrals explicitly, it is more convenient for comparing with the classical time delays to work them into the form\footnote{Equality here requires that $G(\nu_1,\nu_2)$ is analytic in the region of the complex $\nu_1$ plane inside the strip $|\IM\nu_1|\leq\hat a_1$.}
\EQ{
S_{a_1a_2}=\exp\left[\frac{ik  }{\pi}\int^{\nu_1}d\nu_1\,\log\frac{G(\nu_1+i\hat a_1,\nu_2+i\hat a_2)
G(\nu_1-i\hat a_1,\nu_2-i\hat a_2)}{G(\nu_1+i\hat a_1,\nu_2-i\hat a_2)G(\nu_1-i\hat a_1,\nu_2+i\hat a_2)}+\cdots\right]\ .
\label{sxx}
}
Here, the ellipsis represent terms that can only depend on $\nu_2$ and so, given what we said above, can be ignored.

In order to take the semi-classical limit of the S-matrix elements, we must digress to 
consider how to take the semi-classical limit of the 
the dressing phase. The latter is decomposed as
\EQ{
\sigma_{12}=\exp i\big[\chi(x_1^+,x_2^+)-\chi(x_1^-,x_2^+)-\chi(x_1^+,x_2^-)+\chi(x_1^-,x_2^-)\big]\ ,
\label{dcp}
}
where the quantity $\chi(\nu_1,\nu_2)\equiv\chi(x_1=x(\nu_1),x_2=x(\nu_2))$ satisfies a Riemann-Hilbert problem.
As a function of the $\nu_i$, $\chi(\nu_1,\nu_2)$ inherits the branch cuts of $x_i=x(\nu_i)$ corresponding to $\nu_i\in{\EuScript C}$. The Riemann-Hilbert problem takes the 
form
\EQ{
&\chi(\nu_1+\epsilon,\nu_2+\epsilon)+\chi(\nu_1+\epsilon,\nu_2-\epsilon)\\ &+\chi(\nu_1-\epsilon,\nu_2+\epsilon)+\chi(\nu_1-\epsilon,\nu_2-\epsilon)=i\log\Theta(\nu_1,\nu_2)\ ,
\label{mg3}
}
where $\nu_i\in{\EuScript C}$ and $\epsilon$ is an infinitesimal such that $\nu_i\pm \epsilon$ lie on either side of the cut. Note that $x(\nu+\epsilon)=1/x(\nu-\epsilon)$ along the cut. In \cite{Hoare:2011wr} we found that the kernel takes the form\footnote{In \cite{Hoare:2011wr} we used the variable $u=k\nu/\pi$ instead of $\nu$.}
\EQ{
\Theta(\nu_1,\nu_2)=\frac{ \Gamma_{q^2}(1+ik(\nu_1-\nu_2)/\pi)}
{ \Gamma_{q^2}(1-ik(\nu_1-\nu_2)/\pi)}\ ,
}
where the $q$-gamma function satisfies the basic identity
\EQ{
\Gamma_{q^2}(1+x)=\frac{1-q^{2x}}{1-q^2}\Gamma_{q^2}(x)\ .
\label{apa}
}
In \cite{Hoare:2011wr} we provided an integral representation for $\log\Gamma_{q^2}(1+x)$, here we write it as an infinite product of ordinary gamma functions leading to
\EQ{
\log\Theta(\nu_1,\nu_2)=\prod_{j=0}^\infty\frac{\Gamma(1+ik(\nu_1-\nu_2)/\pi+jk)\Gamma(ik(\nu_1-\nu_2)/\pi+(j+1)k)}{\Gamma(-ik(\nu_1-\nu_2)/\pi+(j+1)k)\Gamma(1-ik(\nu_1-\nu_2)/\pi+jk)}\ .
}
The solution to the Riemann-Hilbert problem can be written in terms of a double integral:\EQ{
\chi(x_1,x_2)=i\oint_{|z|=1}\frac{dz}{2\pi i}\,\frac1{z-x_1}\oint_{|z'|=1}\frac{dz'}{2\pi i}\,\frac1{z'-x_2}\log\Theta(\nu(z),\nu(z'))\ .
\label{fdd3}
}

In the semi-classical limit, that is $g\to\infty$ and $k\to\infty$ with $g/k$ fixed, we will find that $\chi(x_1,x_2)$ has an asymptotic expansion of the form $\chi(x_1,x_2)=\sum_{n=0}^\infty \chi^{(n)}(x_1,x_2)g^{1-n}$. Since $x^+-x^-\sim g^{-1}$ means that the dressing phase has leading order behaviour of the form $\log\sigma\sim {\cal O}(g^{-1})$. In this limit, to leading order
\EQ{
\sigma_{12}=\exp\Big[\frac{i\pi g}{k^2}\,\partial_{\nu_1}\partial_{\nu_2}\chi^{(0)}(\nu_1,\nu_2)\Big]\ .
}

Before we take the semi-classical limit, it is useful to first take 
the derivative of the kernel with respect to
$\nu_1$ and $\nu_2$. To leading order in the semi-classical limit
\EQ{
\partial_{\nu_1}\partial_{\nu_2}\log\Theta(\nu_1,\nu_2)=-\frac{2 ik}\pi\coth(\nu_1-\nu_2)+\cdots\ ,
}
Defining
\EQ{
\eta(x_1,x_2)=g\partial_{x_1}\partial_{x_2}\chi^{(0)}(x_1,x_2)\ ,
}
in the semi-classical limit the Riemann-Hilbert problem \eqref{mg3}, when written in terms of the variables $x_i=x(\nu_i)$, becomes
\EQ{
\eta(x_1,x_2)-x_1^{-2}\eta(x_1^{-1},x_2)&-x_2^{-2}\eta(x_1,x_2^{-1})+(x_1x_2)^{-2}\eta(x_1^{-1},x_2^{-1})\\ &=\frac{2k}\pi\cdot
\frac{\coth(\nu(x_1)-\nu(x_2))}{x_1'x_2'}\ .
}
where
\EQ{
x'(\nu)=\frac{2(x+1/x+2(1+\lambda^2)/(1-\lambda^2))}{1-1/x^{2}}\ ,
}
and can be solved uniquely given that $\chi(x_1,x_2)$ is analytic in the region $|x_i|>1$, $i=1,2$:
\EQ{
\eta(x_1,x_2)=\frac{k(1-\lambda^4)}{2\pi}\cdot\frac{x_1-x_2}{x_1x_2((1+\lambda^2)x_1+1-\lambda^2)((1+\lambda^2)x_2+1-\lambda^2)(x_1x_2-1)}\ .
}
From this we then find the leading order behaviour of the dressing phase:
\EQ{
\sigma_{12}=\exp\Big[\frac{2\pi i}{k(1-\lambda^4)}\cdot\frac{((1+\lambda^2)x_1+1-\lambda^2)((1+\lambda^2)x_2+1-\lambda^2)(x_1-x_2)}{(1-x_1^2)(1-x_2^2)(x_1x_2-1)}+\cdots\Big]\ .
}

We can check our result in the AdS$_ 5{\times} S^5$ sigma model limit, $\lambda\to1$. We have
\EQ{
\lim_{\lambda\to1}\eta(x_1,x_2)=g\frac{x_1-x_2}{x_1^2x_2^2(x_1x_2-1)}
}
and integrating twice, we have the known result (e.g.~\cite{Vieira:2010kb}) at leading order in $g^{-1}$,
\EQ{
\lim_{\lambda\to1}\chi^{(0)}(x_1,x_2)=\Big(x_1+\frac1{x_1}-x_2-\frac1{x_2}\Big)\log\Big(1-\frac1{x_1x_2}\Big)\ ,
}
modulo a sum of functions of $x_1$ or $x_2$ individually which do not contribute to the dressing phase $\sigma_{12}$ because of the particular combination in \eqref{dcp}. 

In the sine-Gordon limit $\lambda\to0$, we can write the result as
\EQ{
\lim_{\lambda\to0}\partial_{\nu_1}\partial_{\nu_2}\chi(\nu_1,\nu_2)=\frac\pi{2k}\tanh((\nu_1-\nu_2)/2)+\cdots
}
which corresponds to
\EQ{
\lim_{\lambda\to0}\sigma_{12}=\exp\Big[-\frac{i\pi}{2k}\tanh((\nu_1-\nu_2)/2)\Big]\equiv\exp\Big[-\frac{i\pi}{2k} \tanh(\theta/2)\Big]\ .
\label{pil}
}
In the above, $x_i=x(\nu_i)$ and in the relativistic limit $\nu_i\to\theta_i$ and $\theta=\theta_1-\theta_2$. This latter result can be checked against the explicit expression for the dressing phase in the relativistic theory \cite{Hoare:2011wr}:
\EQ{
\sigma_{12}=\exp\left[-2i\int_0^\infty \frac{dt}t\,\frac{\sinh t\cosh((k-1)t)\sin(2k\theta t/\pi)}{\cosh(kt)\sinh(2kt)}\right]\ ,
\label{fww2}
}
to leading order in the semi-classical limit. Scaling $t\to t/2k$ and then taking $k\to\infty$ gives
\EQ{
\sigma_{12}=\exp\left[-\frac{i}k\int_0^\infty dt\,\frac{\sin(\theta t/\pi)}{\sinh t}+\cdots\right]\ ,
\label{fww3}
}
which can be integrated to give \eqref{pil}.

Using the result established above for the semi-classical limit of the dressing phase we now consider the S-matrix elements themselves. In this case, one finds
\EQ{
\log G^{(1)}&=-2\log(x_1-x_2)+\cdots\ ,\\
\log G^{(2)}&=-\log(x_1-x_2)+\log(1-x_1x_2)+\cdots\ ,\\
\log G^{(3)}&=2\log(1-x_1x_2)+\cdots\ ,
\label{nqq}
}
with $x_i=x(\nu_i)$ and 
where the ellipsis represent terms that depend only on $x_1$ or $x_2$ separately or are of the form $f(x_1)h(x_2)$. These terms are not captured by the Jackiw-Woo formula relating the semi-classical limit of the S-matrix to the classical scattering and so can be discarded.

Notice that the shifted functions in \eqref{sxx} naturally correspond to the $x_i^\pm$ variables for the two bound states: 
\EQ{
x_i^\pm=x(\nu_i\pm i\hat a_i)\ .
}
In addition, once the second term in \eqref{nqq} is integrated it can be written in terms of the energy and momentum of the bound states. This yields the expressions
\EQ{
S^{(1)}_{a_1a_2}&=\exp\Big[\frac{2ik  }{\pi}\int^{\nu_1}d\nu_1\,\log\left|\frac{x_1^+-x_2^-}{x_1^+-x_2^+}\right|^2+\cdots\Big]\ ,\\[5pt]
S^{(2)}_{a_1a_2}&=\exp\Big[\frac{2ik  }{\pi}\int^{\nu_1}d\nu_1\,\log\left|\frac{x_1^+-x_2^-}{x_1^+-x_2^+}\cdot\frac{1-\frac1{x_1^+x_2^+}}{1-\frac1{x_1^+x_2^-}}\right|+\cdots\Big]\ ,\\[5pt]
S^{(3)}_{a_1a_2}&=\exp\Big[\frac{2ik  }{\pi}\int^{\nu_1}d\nu_1\,\log\left|\frac{1-\frac1{x_1^+x_2^+}}{1-\frac1{x_1^+x_2^-}}\right|^2+\cdots\Big]\ ,
}
We can then express $x^\pm_i$ in terms of the kinematic variables $\xi_i$ and change the integral to one over the energy ${\cal E}_1\equiv{\cal E}_1^{\text{(phys)}}$ using the Jacobian
\EQ{
\frac{\partial\nu}{\partial{\cal E}}=\frac{2i\pi\xi\xi^*}{k(\xi-\xi^*)(1-\xi\xi^*)}=-\frac{\pi}{2k\sin\alpha\sinh\theta}\ .
}
where we have taken the semi-classical limit in the last expression.
Finally,
\EQ{
S^{(1)}_{a_1a_2}&=\exp\Big[-i  \int^{{\cal E}_1}\frac{d{\cal E}_1}{\sin\alpha_1\sinh\theta_1}\,
\log\left|\frac{\xi_1-\xi_2^*}{\xi_1-\xi_2}\right|^2+\cdots\Big]\ ,\\[5pt]
S^{(2)}_{a_1a_2}&=\exp\Big[-i  \int^{{\cal E}_1}\frac{d{\cal E}_1}{\sin\alpha_1\sinh\theta_1}\,
\log\left|\frac{\xi_1-\xi_2^*}{\xi_1-\xi_2}\cdot\frac{\xi_1+\xi_2}{\xi_1+\xi_2^*}\right|+\cdots\Big]\ ,\\[5pt]
S^{(3)}_{a_1a_2}&=\exp\Big[-i  \int^{{\cal E}_1}\frac{d{\cal E}_1}{\sin\alpha_1\sinh\theta_1}\,
\log\left|\frac{\xi_1+\xi_2}{\xi_1+\xi_2^*}\right|^2+\cdots\Big]\ .
\label{kpa}
}

Now we make a detailed comparison of the semi-classical limit of the S-matrix and the time delays via the Jackiw-Woo formula \cite{Jackiw:1975im}. The latter results from considering the 
semi-classical interaction of a particle with a potential and gives the S-matrix for the resulting transmission process as
\EQ{
S({\cal E})\thicksim\exp\left[i\int^{\cal E}_{{\cal E}_\text{th}}d{\cal E}'\,\Delta t({\cal E}')\right]\ ,
\label{jw}
}
where ${\cal E}$ is the energy of the particle and $\Delta t({\cal E})$ is the time delay it experiences as it moves through the potential. ${\cal E}_\text{th}$ is the threshold energy. It has been found that this formula can be used to describe the semi-classical limit of the S-matrix of giant magnons in a $1+1$-dimensional QFT with the potential interpreted as a second giant magnon and the particle as the first. Actually what is successfully captured is the part of the S-matrix that depends non-trivially on the rapidities of both states. However, recall that there are pieces of the quantum S-matrix that depend on the rapidities of the individual particles which are not captured by the Jackiw-Woo formula and which we have not kept track of.

First of all we have to match up the quasi-classical states with the classical giant magnons. The former are determined by taking the $a$-fold product of the basic states \eqref{sdp} whereas as the latter are determined by the internal collective coordinates as in \eqref{xff}. It is clear that the  collective coordinates and coherent states match precisely as we expect on the basis of the correspondence principle. 

Now we can compare individual processes. Matching \eqref{num1} to \eqref{wea2}, the relation of S-matrix elements to the parameters $c_i$ and $c_i'$ of the magnons is
\EQ{
&S^{(1)}:\qquad c_1=c'_1=1\ ,\quad c_2=c'_2=0\ ,\\
&S^{(2)}:\qquad c_1=c'_2=1\ ,\quad c_2=c'_1=0\ ,\\
&S^{(3)}:\qquad c_2=c'_2=1\ ,\quad c_1=c'_1=0\ ,\\
}
It is then straightforward to see that the Jackiw-Woo formula \eqref{jw} yields precisely the expressions \eqref{kpa} for the semi-classical S-matrix elements.

\section{The XXZ spin chain connection}\label{s6}

In the sigma model limit, $k\to\infty$ with $g$ fixed, the dual gauge theory is weakly coupled when $g$ is small. In this limit, the magnons can be mapped on to the {\it magnon\/} excitations of a spin chain. In the $\msu(2)$ sector and at one loop order $g^2$ this is simply the Heisenberg XXX spin chain \cite{Beisert:2003yb,Beisert:2004ry,Minahan:2010js}. In this limit, the magnon dispersion relation is directly related to the energy of excitations of the spin chain in its thermodynamic limit \cite{Dorey:2006dq}. Firstly, from \eqref{yr2} we find
\EQ{
{\cal E}=a+\frac{4g^2}{a}\Big(1-\cos\frac{\cal P}{2g}\Big)+{\cal O}(g^4)\ .
}
The energy of the spin chain is then related to this by a simply addition and scaling:
\EQ{
E_\text{s.c.}=\frac1{2g^2}\big({\cal E}-a\big)=\frac4{a}\sin^2\frac{K}2
}
where $K={\cal P}/2g$ is the momentum of the spin chain. The energy above is the well known energy of a bound state of $a$ basic magnons of the spin chain in the thermodynamic limit.

The $\msu(2)$ sector of the Heisenberg spin chain describes operators in the dual gauge theory that are single trace
of length $L$, the length of the chain, built form two of the complex scalars $X$ and $Y$. The ferromagnetic vacuum of the chain $\ket{\uparrow\uparrow\cdots\uparrow}$ corresponds to the operator $\Tr(X^L)$ while a state with one down spin $\ket{\uparrow\cdots\uparrow\downarrow\uparrow\cdots}$ to $\Tr(X\cdots XYX\cdots)$. 

In the lambda model it is also interesting to consider the $g\to0$ limit of the magnon dispersion relation. In this limit, the mass shell condition can be solved as a series in $g$:
\EQ{
{\cal E}=a+\frac{4\pi g^2}{k\sin(\pi a/k)}\Big(\cos\frac{\pi a}{k}-\cos\frac{\cal P}{2g}\Big)+{\cal O}(g^4)\ .
\label{ens}
}

Remarkably this dispersion relation is precisely that of the XXZ spin chain in its paramagnetic regime. Let us digress to explain this in more detail. The XXZ spin chain has a Hamiltonian of the form\footnote{Our discussion of the XXZ spin chain draws on the book \cite{Tak} and the review article \cite{Samaj}.}
\EQ{
H_\text{s.c.}=-\frac12\sum_n\big(\sigma^x_n\sigma^x_{n+1}+\sigma^y_n\sigma^y_{n+1}+\Delta(\sigma^z_n\sigma^z_{n+1}-1)\big)
}
where in our case the relevant value of $\Delta$ is
\EQ{
\Delta=-\cos\gamma=\cos\frac\pi k\ ,\qquad \gamma=\pi\frac{k-1}k\ .
\label{wiz}
}
The XXX spin chain is recovered, as it should be, in the limit $k\to\infty$.

The ground state of the XXZ spin chain when $\Delta<1$ is no longer the ferromagnetic ground state with all spins up $\ket{\uparrow\uparrow\cdots\uparrow}$. However, this state does provide a perfectly good reference state for the coordinate Bethe ansatz. An eigenstate with $M$ spins down has an energy given by
\EQ{
E_\text{s.c.}=2\sum_{i=1}^M(\Delta-\cos K_i)\ ,
}
where the wave numbers are determined by the Bethe equations defined as follows. Firstly define another rapidity $\eta$ via
\EQ{
e^{iK}=\frac{\sinh\frac12(i\gamma-\eta)}{\sinh\frac12(i\gamma+\eta)}\ ,
}
then the allowed rapidities satisfy the Bethe ansatz equations
\EQ{
\left[\frac{\sinh\frac12(i\gamma-\eta_i)}{\sinh\frac12(i\gamma+\eta_i)}\right]^L=
\prod_{j(\neq i)}^M\frac{\sinh\frac12(\eta_i-\eta_j-2i\gamma)}{\sinh\frac12(\eta_i-\eta_j+2i\gamma)}\ .
\label{bae}
}

The solution of these equations simplifies considerably in the limit of a very long chain $L\to\infty$, at least for states with a finite number of down spins $M$. The solutions to the Bethe ansatz equations in the $L\to\infty$ limit come in the form of two kinds of strings: positive ``parity" with
\EQ{
\eta_j=\eta+i\gamma(M+1-2j)\ ,\qquad j=1,2,\ldots,M\ ,
\label{seven}
}
and negative ``parity" with
\EQ{
\eta_j=\eta+i\pi+i\gamma(M+1-2j)\ ,\qquad j=1,2,\ldots,M\ .
\label{sodd}
}

However, there are constraints on these strings that arise from the normalizability of the associated state. For our particular choice of $\gamma$ in \eqref{wiz} these selection rules require states with $M$ even to have even parity and states with $M$ odd to have odd parity. In addition, with $\gamma=\p(k-1)/k$, the length of the strings is restricted to be $M<k$. 

So rather serendipitously, the bound states are naturally restricted in a way that meshes with the quantum group representation theory with $q$ a $2k$ root of unity. This point deserves amplifying. The $\msu(2)$ which acts naturally on the spin chain is {\it not\/} a subgroup of the stability group $S$ of $\Lambda$ and so is not subject to a $q$ deformation and IRF/RSOS restriction. However, the basic magnon of the spin chain, the $\ket{\downarrow}$ state, corresponding to the $Y$ field insertion in the dual gauge theory, can be chosen to carry $(m_1=\frac12,m_2=\frac12)$ quantum numbers under $\SU(2)_3{\times}\SU(2)_4\subset\SU(4)$. So the $a$ (assumed to be positive) bound state in the spin chain has $(m_1=a/2,m_2=a/2)$ and so lies in a multiplet with spin $(j_1=a/2,j_2=a/2)$. The quantum group restriction imposes the condition $a<k-1$, matching closely the kinematical restriction on $a$ from the spin chain. 

Summing up the energies of all the constituents of a string reveals that they have an energy 
\EQ{
E_\text{s.c.}=\frac{2\sin(\pi/k)}{\sin(\pi M/k)}\Big(\cos\frac{\pi M}k+(-1)^M\cos K\Big)\ .
}
The normalizability of the bound state imposes conditions that restrict the momenta to lie in the ranges
\EQ{
(M\text{ even}):\qquad |K|<\pi-\frac{\pi M}k\ ,\qquad (M\text{ odd}):\qquad \pi>|K|>\frac{\pi M}k\ .
\label{mcd}
}
In the limit, $k\to\infty$, the strings match the strings of the XXX spin chain and the XXX momentum of the bound states is $K$, for odd parity strings, and $\pi-K$, for even parity strings.

The relation to our magnon dispersion relation in the $g\to0$ limit \eqref{ens} now reveals itself. The spin chain energy of a $M$ bound state is related to the energy of the $a=M$ magnons in the lambda theory via
\EQ{
E_\text{s.c.}=\frac{k\sin(\pi/k)}{2\pi g^2}({\cal E}-a)\ .
}
In addition, the momenta are related via
\EQ{
(a=M\text{ even}):\qquad K=\pi-\frac{{\cal P}}{2g}\ ,\qquad (a=M\text{ odd}):\qquad K=\frac{\cal P}{2g}\ .
}
The condition on the momentum of a bound state \eqref{mcd}, is interesting because it corresponds precisely to the magnon branch identified in section \ref{s2}, that is for $g\to0$
\EQ{
2\pi g>|{\cal P}|>2\pi g\frac{a}k\ ,
}

In closing it is worth pointing out that the spin chain rapidity $\eta$ is related in a simply way to the pseudo rapidity $\nu$ that we defined in terms of $x^\pm$ in \eqref{j22} and \eqref{j23}:
\EQ{
\eta=i\pi-2\nu\ (\text{mod }2\pi i)\ .
}
The strings \eqref{seven} and \eqref{sodd} are then equal precisely to the bound state strings \eqref{etas} of the S matrix.

The relation to the XXZ spin chain is deeper than just an equivalence of mass shell conditions for magnon bound states. The S-matrix itself, in the $\msu(2)$ sector, actually reduces to the XXZ spin chain S-matrix as it should. The scattering of two basic magnon states in the $\msu(2)$ sector is the  element $S^{(1)}$ in \eqref{xpp}. Now we take the limit of this element as $g\to0$. In this limit,
\EQ{
x^\pm\longrightarrow\frac k{4\pi g}\big(1+q^{\pm1}e^{2\nu}\big)+{\cal O}(g^0)
}
and then the S matrix element can be written in terms of the pseudo rapidity difference $\nu=\nu_1-\nu_2$ on the branch with $\alpha>0$. It takes the simple form
\EQ{
S^{(1)}\longrightarrow\frac{\sinh(\nu-\frac{\pi i}k)}{\sinh(\nu+\frac{\pi i}k)}=\frac{\sinh\frac12(\eta-2i\gamma)}{\sinh\frac12(\eta+2i\gamma)}\ .
}
But this is precisely the scattering phase of the XXZ model as is evident from the Bethe ansatz equations \eqref{bae}.

The details of how to generalize this to the magnon bound state multiplets will be presented elsewhere. It seems tempting to think that there is a relation between the spin chain that describes the lambda model at small $g$ and the quantum group deformation of the spin chain of the ${\cal N}=4$ theory proposed in \cite{Berenstein:2004ys}. However, there are puzzles to understand; for instance, how to integrate the properties of the representation theory of quantum groups for $q$ a root of unity with the spin chain.

\section*{Acknowledgements}

CA and DP are supported by STFC studentships.
TJH is supported in part by the STFC grant ST/G000506/1. 
JLM is supported in part by MINECO-Spain (FPA2014-52218-P), Xunta de Galicia-Conseller\'\i a de Educaci\'on (AGRUP2015/11), and FEDER.
DMS was supported in part by the post-doc FAPESP grant 2012/09180-9.

\appendix
\appendixpage

\section{Conserved charges}\label{a1}

In this appendix, we briefly review how to extract an infinite series of conserved currents and associated charges for the lambda or sigma model from the Lax connection. We refer to the book \cite{BBT} for the general analysis. However, there are some features that are special to the present context that are worth spelling out.  

The starting point are the Lax equations \eqref{leq2}. Conserved currents can be constructed by the process of {\it abelianization\/} around either the pole at $z=0$ or $z=\infty$ of the Lax connection \eqref{mm2b}. Let us take the pole at $z=\infty$, although there is an analogous analysis for the pole at $0$. In the gauge fixed theory, the Lax connection takes the form~\eqref{gfLax} and the component multiplying the double pole $A_+^{(2)}$ lies on the $G$ adjoint orbit of the constant element $\Lambda$. The idea then is to construct a (local) gauge transformation $\Phi(z)=\sum_{n=0}^\infty\Phi_{-n}z^{-n}$ of the Lax connection order-by-order in $z$ such that it takes values in the subalgebra $\mathfrak s\subset\mathfrak f$ that commutes $\Lambda$. In a generic case, this would be a Cartan subalgebra---hence the name {\it abelianization\/}---however, here $\Lambda$ is not regular and in our case $\mathfrak s$ is non abelian. In fact $\mathfrak s=\mathfrak{ps}\big(\mathfrak{u}(2|2)\oplus\mathfrak{u}(2|2)\big)$. For the gauge transformed Lax connection $\AA'_\pm(z)$, the component $\AA_+^{(2)\prime}=\Lambda$.
The Lax equation still takes the form
\EQ{
[\partial_\mu+\AA'_\mu(z),\partial_\nu+\AA'_\nu(z)]=0\ ,
\label{leq3}
}
however, if we take the super trace of this equation with $\Lambda$ then the commutator term vanishes because $\AA'_\mu\in\mathfrak s$. It follows that there exist local conserved currents generated by 
\EQ{
\EuScript J_\pm(z)=\pm\STr\big[\Lambda \AA'_\pm(z)]\ .
}

We can then define a further (non local) gauge transformation $\Omega(z)=1+\sum_{n=1}^\infty\Omega_{-n}z^{-n}$ in the group generated by $\mathfrak s$ in order to transform the connection into
\EQ{
\AA''_\pm(z)=z^{\pm2}\Lambda\ .
}
If we define the two gauge transformations together as $\Theta_-(z)=\Phi(z)\Omega(z)$, then it becomes apparent that this can be interpreted as a dressing transformation and the wave function takes the form  \eqref{jff}
\EQ{
\Psi(z)=\Phi(z)\Omega(z)\Psi_0(z)g_-(z)=\Theta_-(z)\Psi_0(z)g_-(z)\ .
}

The important point is that the gauge transformation $\Phi(z)$ is local in the underlying fields of the gauge fixed worldsheet theory. As a consequence the gauge transformation is periodic on the world sheet $\Phi(\pi;z)=\Phi(-\pi;z)$. It follows that the ``right monodromy" defined in \eqref{mon} equals
\EQ{
{\cal W}(z)=\Omega(-\pi;z)^{-1}\Omega(\pi;z)\ .
}
Since $\Omega(x;z)$ lies in the group that stabilizes $\Lambda$ it follows that
\EQ{
\STr\big[\Lambda\log{\cal W}(z)\big]=\STr\big[\Lambda\log T'(z)\big]
\label{klo}
}
where $T'(z)=\Omega(\pi;z)\Omega(-\pi;z)^{-1}$ is the monodromy (Wilson line) of the gauge transformed Lax connection $\AA'_\mu(z)$:
\EQ{
T'(z)=\overset{\longleftarrow}{\text{Pexp}}\Big[-\int_{-\pi}^\pi d\sigma\,\AA'_1(\sigma;z)\Big]\ .
}
Next, we remark that since $\AA'(\sigma;z)$ commutes with $\Lambda$ we can remove the path ordering in \eqref{klo}
\EQ{
\STr\big[\Lambda\log T'(z)\big]&=-\STr\Big[\Lambda\int_{-\pi}^\pi d\sigma\,\AA'_1(\sigma;z)\Big]\\ &=\int_{-\pi}^\pi d\sigma\,\EuScript J_0(\sigma;z)\equiv\mathfrak Q(z). 
}
Hence, it follows that the conserved charges associated to the local conserved currents $\EuScript J_\mu(z)$ can be expressed as
\EQ{
\mathfrak Q(z)=\STr\big[\Lambda\log{\cal W}(z)\big]\ .
} 

It is worth pointing out that in the HM limit, where the spatial coordinate runs from $-\infty$ to $+\infty$, for field configurations that approach the vacuum at $\pm\infty$, i.e.~$\Phi(\pm\infty;z)=1$, the charge $\mathfrak Q(z)$ can also be expressed in terms of the monodromy (Wilson line) of the original Lax connection $T(z)$ \cite{Hollowood:2011fq,Hollowood:2010dt}:
\EQ{
{\mathfrak Q}(z)=\STr\big[\Lambda\log{\cal W}(z)\big]=\STr\big[\Lambda\log T(z)\big]\ .
}

\section{Noether symmetries}\label{a2}

In this appendix, we show that the Noether charges $\mathfrak Q(\lambda^{\pm1/2})$ generate symmetries of the Lagrangian of the lambda model. It is possible to show this in the theory before gauge fixing, but the discussion is complicated. It is much simpler to discuss the symmetries in the gauge fixed theory and we will satisfy ourselves with this.

If we vary the field $\CF$, the variation of the action can be written in two equivalent ways as
\EQ{
\delta S&=\frac k{2\pi}\int d^2x\,\STr\Big(\delta\CF \CF^{-1}[\partial_++\AA_+(\lambda^{1/2}),\partial_-+\AA_-(\lambda^{1/2})]\Big)\\ &=\frac k{2\pi}\int d^2x\,\STr\Big(\CF^{-1}\delta\CF[\partial_++\AA_+(\lambda^{-1/2}),\partial_-+\AA_-(\lambda^{-1/2})]\Big)\ .
\label{xix}
}
Note that the special values of the spectral parameter $z=\lambda^{\pm1/2}$ appear quite naturally.

Let us pick the second expression in \eqref{xix}. We can use the gauge transformation $\Phi(\lambda^{-1/2})$, defined in appendix \ref{a1}, local in the field, to gauge transform the Lax connection to $\AA'_\mu(\lambda^{-1/2})$. To this end, we define the variation
\EQ{
\delta\CF=\CF\Phi(\lambda^{-1/2})^{-1}\Lambda\Phi(\lambda^{-1/2})\ ,
}
then
\EQ{
\delta S=-\frac k{4\pi}\int d^2x\,\partial^\mu{\EuScript J}_\mu(\lambda^{-1/2})=0\ ,
}
on shell.

Similarly, for the first expression in \eqref{xix}, with
\EQ{
\delta\CF=\Phi(\lambda^{1/2})^{-1}\Lambda\Phi(\lambda^{1/2})\CF\ ,
}
gives
\EQ{
\delta S=-\frac k{4\pi}\int d^2x\,\partial^\mu{\EuScript J}_\mu(\lambda^{1/2})=0\ ,
}
on shell.
So this identifies ${\EuScript J}_\mu(\lambda^{\pm1/2})$ as the currents associated to symmetries of the Lagrangian and ${\mathfrak Q}(\lambda^{\pm1/2})$ are the corresponding Noether charges.   

We now show that in the limit $\lambda\to1$
\EQ{
\lim_{\lambda\to1}\frac1{\lambda-\lambda^{-1}}\big[\EuScript J_\mu(\lambda^{-1/2})-\EuScript J_\mu(\lambda^{1/2})\big]\ ,
\label{rdo}
}
becomes the Noether current for left $F$ transformations in the sigma model of the form $\delta f=\Lambda f$.
For a general $F_L$ transformation $f\to Uf$, the conserved current takes the form
\EQ{
J^L_\pm=f\Big(\pm\frac12(f^{-1}\partial_\pm f)^{(1)}+(f^{-1}\partial_\pm f)^{(2)}\mp\frac12(f^{-1}\partial_\pm f)^{(3)}\Big)f^{-1}\ .
}
This current is related to the Lax connection by expanding around the point $z=1$; defining $z=1+\varepsilon$,
\EQ{
\partial_\pm+\AA_\pm(1+\varepsilon)=f^{-1}\big(\partial_\pm\pm2\varepsilon J^L_\pm\big)f+{\cal O}(\varepsilon^2)\ .
}
This shows that the Noether current for the transformation $\delta f=\Lambda f$, is precisely given by the limit \eqref{rdo}.

\section{Symplectic form}\label{a3}

The symplectic form of the lambda model can be constructed in a covariant way directly from the Lagrangian (see for example \cite{Witten:1986qs,Aldaya:1991qw}) 
\EQ{
\omega=\int_\Sigma d\Sigma_\mu\,S^\mu\ ,
}
for a suitable Cauchy surface $\Sigma$,
where the symplectic current takes the form\footnote{The expression in \eqref{pp2} is valid for a theory with  an action at most quadratic in derivatives. For the general case,
\EQ{
\delta L=\partial_\mu j^\mu+\text{EoM}_a\delta\phi_a\ ,\qquad S^\mu=-\delta j^\mu\ .
}
So on-shell $\delta L=\partial_\mu j^\mu$. Hence, using $\delta^2=0$, we have $\delta^2L=0=\partial_\mu \delta j^\mu=-\partial_\mu S^\mu$ and then trivially $\delta S^\mu=0$.}
\EQ{
S^\mu=\delta\phi_a\wedge \delta\frac{\partial L}{\partial \partial_\mu\phi_a}\ ,\qquad \partial_\mu S^\mu=0\ ,
\label{pp2}
}
or in terms of forms:
\EQ{
\omega=\int_\Sigma *S\ ,\qquad S=S_\mu dx^\mu\ .
}
Note that $\omega$ is closed and does not depend on the choice of Cauchy surface $\Sigma$ precisely because $S^\mu$ is conserved. 

In the lambda model, we find that the symplectic current has components
\EQ{
S_-&=-\frac k{4\pi}\STr\big(\delta \CF  \CF ^{-1}\wedge\delta(\partial_-\CF \CF ^{-1})-2\CF ^{-1}\delta \CF \wedge \CF ^{-1}\delta \CF A_--2\CF ^{-1}\delta \CF \wedge\delta A_-\big)\ ,\\
S_+&=-\frac k{4\pi}\STr\big(\CF ^{-1}\delta \CF  \wedge\delta(\CF ^{-1}\partial_+\CF )-2\delta \CF  \CF ^{-1}\wedge \delta \CF  \CF ^{-1}A_++2\delta \CF \CF ^{-1}\wedge\delta A_+\big)\ .
\label{pb2}
}
These are, of course, precisely the components of the symplectic current in the gauged WZW model since the deformation does not affect the kinetic terms.

It is useful to write the symplectic form in terms of the wave function defined at the two special points $z=\lambda^{\pm1/2}$
\EQ{
\Psi_{{\sst(\pm)}}(x)\equiv\Psi(x;\lambda^{\pm1/2})\ .
}
In terms of these quantities
\EQ{
\CF=\Psi_{{\sst(+)}}\Psi_{{\sst(-)}}^{-1}\ ,\qquad A_\pm=-\partial_\pm\Psi_{{\sst(\pm)}}\Psi_{{\sst(\pm)}}^{-1}\ .
}
It then follows that
\EQ{
\delta A_\pm&=-\Psi_{{\sst(\pm)}}\partial_\pm\big(\Psi_{{\sst(\pm)}}^{-1}\delta\Psi_{{\sst(\pm)}}\big)\Psi_{{\sst(\pm)}}^{-1}\ ,
}
as well as
\EQ{
&\delta\CF\CF^{-1}=\Psi_{{\sst(+)}}\big(\Psi_{{\sst(+)}}^{-1}\delta\Psi_{{\sst(+)}}-\Psi_{{\sst(-)}}^{-1}\delta\Psi_{{\sst(-)}}\big)\Psi_{{\sst(+)}}^{-1}\ ,\\
&\CF^{-1}\delta\CF=\Psi_{{\sst(-)}}\big(\Psi_{{\sst(+)}}^{-1}\delta\Psi_{{\sst(+)}}-\Psi_{{\sst(-)}}^{-1}\delta\Psi_{{\sst(-)}}\big)\Psi_{{\sst(-)}}^{-1}\ .
}

One then finds that
\EQ{
S_0&=-\frac k{4\pi}\STr\Big[\Psi_{{\sst(-)}}^{-1}\delta\Psi_{{\sst(-)}}\wedge\partial_1(\Psi_{{\sst(-)}}^{-1}\delta\Psi_{{\sst(-)}})
-\Psi_{{\sst(+)}}^{-1}\delta\Psi_{{\sst(+)}}\wedge\partial_1(\Psi_{{\sst(+)}}^{-1}\delta\Psi_{{\sst(+)}})\\ &\qquad\qquad\qquad+\partial_1\big(\Psi_{{\sst(-)}}^{-1}\delta\Psi_{{\sst(-)}}\wedge\Psi_{{\sst(+)}}^{-1}\delta\Psi_{{\sst(+)}}\big)\Big]\ .
\label{bxx}
}
When we integrate this around the world sheet to find the symplectic form, the final term contributes at the boundaries $\sigma=\pm\pi$. It is useful to write
\EQ{
\omega=\omega_{{\sst(+)}}-\omega_{{\sst(-)}}\ ,
\label{ju1}
}
where
\EQ{
\omega_{{\sst(\pm)}}&=\frac k{4\pi}\int_{-\pi}^\pi d\sigma\,\STr\Big[\Psi_{{\sst(\pm)}}^{-1}\delta\Psi_{{\sst(\pm)}}\wedge\partial_1(\Psi_{{\sst(\pm)}}^{-1}\delta\Psi_{{\sst(\pm)}})\Big]\\ &\qquad\qquad+
\frac k{4\pi}\STr\Big[\Psi_{{\sst(\pm)}}(-\pi)^{-1}\delta\Psi_{{\sst(\pm)}}(-\pi)\wedge\delta{\cal W}{\cal W}^{-1}\Big]\ .
\label{ju2}
}
One might recognize $\omega_{{\sst(\pm)}}$ as the symplectic forms of the chiral WZW model. 
In that context, the WZW field $g=g_1(x^+)g_2(x^-)^{-1}$ and the symplectic form is as above with $\Psi_{{\sst(+)}}\sim g_1$ and $\Psi_{{\sst(-)}}\sim g_2$. This is rather remarkable because in the present circumstances the split $\CF=\Psi_{{\sst(+)}}\Psi_{{\sst(-)}}^{-1}$ is not a chiral split in terms of the world sheet coordinates.

Note that although the symplectic form $\omega$ is closed, the components $\omega_{{\sst(\pm)}}$ are not separately closed due to the boundary term; in fact
\EQ{
\delta\omega_{{\sst(\pm)}}&=\frac k{12\pi}\STr\big[{\cal W}^{-1}\delta{\cal W}\wedge{\cal W}^{-1}\delta{\cal W}\wedge{\cal W}^{-1}\delta{\cal W}\big]\ .
\label{peq}
}

We can write the symplectic form in a rather elegant way by introducing a twisted  inner product on the loop group
of $F$ \cite{Delduc:2013fga,Vicedo:2015pna}
\EQ{
\big\langle a,b\big\rangle_{\phi }=\oint\frac{dz}{2\pi iz}\phi(z)\STr\big(a(z)b(z)\big)\ ,
\label{inner}
}
with twist function
\EQ{
\phi (z)=\frac k{2\pi}\cdot\frac{\lambda^2-\lambda^{-2}}{z^4-\lambda^2-\lambda^{-2}+z^{-4}}\ .
\label{twf}
}

In terms of this inner product, we can write the symplectic form as 
\EQ{
\omega&=\frac 1{2}\int_{-\pi}^\pi d\sigma\,\big\langle\Psi^{-1}\delta\Psi,\wedge\partial_1(\Psi^{-1}\delta\Psi)\big\rangle_{\phi}+
\frac 1{2}\big\langle\Psi(-\pi)^{-1}\delta\Psi(-\pi),\wedge\delta{\cal W}{\cal W}^{-1}\big\rangle_\phi\ .
\label{ju3}
}

We can also write the symplectic form in terms of the Kac-Moody currents $\JJ_\mu$ defined in~\cite{Hollowood:2014qma}. The latter are related to the wave function via 
\EQ{
\JJ_\pm=\pm\frac k{2\pi} \partial_1\Psi_{\sst(\mp)}\Psi_{\sst(\mp)}^{-1}
}
and so it follows that
\EQ{
\delta\!\JJ_\pm=\big[\pm\frac k{2\pi}\partial_1-\JJ_\pm,\delta\Psi_{\sst(\mp)}\Psi_{\sst(\mp)}^{-1}\big]
}
and \eqref{bxx} leads to
\EQ{
\omega=\frac{2\pi}k\int_{-\pi}^\pi d\sigma\,\STr\big(-\delta\!\JJ_+\big[\partial_1-\frac{2\pi}k\text{ad}\JJ_+\big]^{-1}\delta\!\JJ_++\delta\!\JJ_-\big[\partial_1+\frac{2\pi}k\text{ad}\JJ_-\big]^{-1}\delta\!\JJ_-\big)\ .
\label{bxx3}
}

Inverting the symplectic form gives the Poisson bracket algebra of the Kac-Moody currents,
\EQ{
\big\{\JJ^a_\pm(\sigma),\JJ^b_\pm(\sigma')\big\}&=f^{ab}{}_c\JJ_\pm^c(\sigma')\delta(\sigma-\sigma')\mp\frac {k}{2\pi}\eta^{ab}\delta'(\sigma-\sigma')\ ,\\
\big\{\JJ^a_+(\sigma),\JJ^b_-(\sigma')\big\}&=0\ ,
\label{km4}
}
with $\JJ_\pm^a=\STr(T^a\JJ_\pm)$. So the Poisson brackets of the lambda model can be written, as above, in a lambda-independent way as two classical commuting Kac-Moody algebras \cite{Hollowood:2014rla,Hollowood:2014qma}.

\end{document}